\documentclass[aps,graphicx,10pt,showkeys,showpacs,twocolumn]{revtex4}
\usepackage{amsmath}
\usepackage{amscd}
\usepackage{graphicx}
\usepackage{subfigure}
\usepackage{appendix}
\usepackage{multirow}
\usepackage{mathrsfs}

\begin{document}
\title[Short Title]{Extension of Noether's theorem in $\mathcal{PT}$-symmetric systems and its experimental demonstration in an optical setup}

\author{Qi-Cheng Wu$^{1,}$\footnote{These authors contributed equally to this work}}
\author{Jun-Long Zhao$^{1,*}$}
\author{Yu-Liang Fang$^{1}$}
\author{Yu Zhang$^{1,2}$}
\author{Dong-Xu Chen$^{1,}$\footnote{E-mail: chendx@sru.edu.cn}}
\author{Chui-Ping Yang$^{1,3,}$\footnote{E-mail: yangcp@hznu.edu.cn}}
\author{Franco Nori$^{4,5,6,}$\footnote{E-mail: fnori@riken.jp}}

\affiliation{$^{1}$Quantum Information Research Center, Shangrao Normal University, Shangrao 334001, China\\
             $^{2}$School of Physics, Nanjing University, Nanjing 210093, China\\
                        $^{3}$Hangzhou Normal University, Hangzhou, Zhejiang 311121, China\\
             $^{4}$Theoretical Quantum Physics Laboratory, RIKEN, Wako-shi, Saitama 351-0198, Japan\\
             $^{5}$RIKEN Center for Quantum Computing, RIKEN, Wako-shi, Saitama 351-0198, Japan\\
             $^{6}$Physics Department, The University of Michigan, Ann Arbor, Michigan 48109-1040, USA}

\begin{abstract}
Noether's theorem is one of the fundamental laws in physics,
relating the symmetry of a physical system to its constant of
motion and conservation law. On the other hand, there exist a
variety of non-Hermitian parity-time ($\mathcal{PT}$)-symmetric
systems, which exhibit novel quantum properties and have attracted
increasing interest. In this work, we extend Noether's theorem to
a class of significant $\mathcal{PT}$-symmetric systems for which
the eigenvalues of the $\mathcal{PT}$-symmetric Hamiltonian
$\hat{H}_\mathcal{PT}$  change from purely real numbers to purely
imaginary numbers, and introduce a generalized expectation value
of an operator based on biorthogonal quantum mechanics. We find
that the generalized expectation value of a time-independent
operator is a constant of motion when the operator presents a
standard symmetry in the $\mathcal{PT}$-symmetry unbroken regime,
or a chiral symmetry  in the $\mathcal{PT}$-symmetry  broken
regime. In addition, we experimentally investigate the extended
Noether's theorem in $\mathcal{PT}$-symmetric single-qubit and
two-qubit systems using an optical setup. Our experiment
demonstrates the existence of the constant of motion and reveals
how this constant of motion can be used to judge whether the
$\mathcal{PT}$-symmetry of a system is broken. Furthermore, a
novel phenomenon of masking quantum information is first observed
in a $\mathcal{PT}$-symmetric two-qubit system. This study not
only contributes to full understanding of the relation between
symmetry and conservation law in $\mathcal{PT}$-symmetric physics,
but also has potential applications in  quantum information theory
and quantum communication protocols.
\end{abstract}



\maketitle

\footnotesize


\section{Introduction}
The subject of finding the symmetries of dynamics is of
fundamental interest and has broad applications in physics, e.g.,
high-energy scattering experiments, control issues in mesoscopic
physics and quantum
cosmology~\cite{Symmetry1,chiralsymmetry1,chiralsymmetry2,chiralsymmetry3,chiralsymmetry4,topological2}.
On the other hand, by means of symmetries, one can generally make
non-trivial inferences from complex systems, such as many-body
systems, dissipative systems and non-Hermitian systems.  As
an important theorem which is related to symmetries, Noether's
theorem~\cite{Noether1} has important applications in quantum
physics and quantum information
science~\cite{Noether2,Ehrenfest1,Ehrenfest2,Ehrenfest3,Noetheranalysis,Noethercurrents}.
Noether's theorem states that every symmetry of dynamics implies a
conservation law, and it was originally applied in Lagrangian
approach in classical mechanics to uncover conserved quantities
from symmetries of the Lagrangian. In many cases, the existence of
these conserved quantities is very important for understanding the
physical states and the properties of the
systems~\cite{Noether2,Ehrenfest2,Noetheranalysis,Noethercurrents}.
The theorem applies also in  quantum mechanics, and the most
prominent example of Noether's theorem is Ehrenfest's theorem in
closed systems~\cite{Ehrenfest1,Ehrenfest}
\begin{eqnarray}\label{eqm-1}
    \frac{d}{dt}\langle{\hat{F}}\rangle=\frac{1}{i\hbar}\langle{[\hat{F},\hat{H}]}\rangle+\langle{\frac{d\hat{F}}{dt}}\rangle.
\end{eqnarray}
For an operator $\hat{F}$ without explicit time dependence, it
then follows that its expectation value $\langle{\hat{F}}\rangle$
is a constant of motion if it commutes with the Hermitian
Hamiltonian $\hat{H}$. However, Ehrenfest's theorem is not
applicable for open
systems~\cite{Ehrenfest3,Ehrenfest,open1,r2,r5}. Furthermore, even
in  closed systems,  Ehrenfest's conservation law cannot capture
all features of symmetry when mixed states are
considered~\cite{Ehrenfest2}.

A natural extension of Noether's theorem in non-Hermitian systems
is to replace the Hermitian Hamiltonian $\hat{H}$ with a
non-Hermitian Hamiltonian $\hat{H}^\dag$, which turns
eq.~(\ref{eqm-1}) into
${d\langle{\hat{F}}\rangle}/{dt}=\frac{1}{i\hbar}\langle{[\hat{F}\hat{H}-\hat{H}^\dag{\hat{F}}]}\rangle
+\langle{{d\hat{F}}/{dt}}\rangle$~\cite{conservationlaws2,interwining1,interwining2,SRM}.
Up to now, based on the important intertwining relation
$\hat{F}\hat{H}=\hat{H}^\dag{\hat{F}}$~\cite{interwining1,interwining2,SRM},
several methods have been proposed to obtain conserved quantities,
including spectral decomposition methods~\cite{SDM1,SDM2},
recursive construction of intertwining operators~\cite{RCM},
sum-rules method~\cite{SRM},  Stokes parametrization
approach~\cite{SPA}, and so on. Recently, the authors in
ref.~\cite{Ehrenfest4}  investigated a manifestation of Noether's
theorem in non-Hermitian systems, where an inner product was
defined as $(\varphi,\psi)\equiv\varphi_{u}^{T}\psi_{v}$ without
its complex conjugation. In their framework, a generalized
symmetry, which they termed pseudochirality, emerges naturally as
the counterpart of the symmetry defined by the commutation
relation in quantum mechanics. Some existing
works~\cite{Ehrenfest1,Ehrenfest2,Ehrenfest3,Ehrenfest,Ehrenfest4,Noetheranalysis,Noethercurrents,conservationlaws2,open1,interwining1,interwining2,SRM,SDM1,SDM2,RCM,SPA,r2,r5}
enrich the understanding of obtaining conserved quantity beyond
the Hermitian framework, whereas a full understanding of the
relation between symmetry and conservation law, and practical
methods for extracting expectation values in non-Hermitian
systems, remain elusive. Therefore, in order to properly deal with
conservation problems using Noether's theorem and explore its
potential applications in non-Hermitian systems, there is an
urgent need to extend Noether's theorem to non-Hermitian systems.

Over the past decades, there is considerable interest in the study
of the dynamic properties of parity-time
($\mathcal{PT}$)-symmetric non-Hermitian
systems~\cite{Non-Hermitian1,Non-Hermitian2,Non-Hermitian3,Non-Hermitian4,Non-Hermitian5,Non-Hermitianadd,r1,r6}.
The  unique properties of $\mathcal{PT}$-symmetric systems and
their applications have been investigated in various physical
systems~\cite{topological1,Non-Hermitian6,optomechanics1,optomechanics2,photonics1,photonics2,microwave1,
microwave2,r3,r4}. Moreover, many remarkable and unexpected
quantum phenomena have been observed in $\mathcal{PT}$-symmetric
systems, such as critical
phenomena~\cite{CriticalPhenomena1,CriticalPhenomena2}, chiral
population transfer~\cite{energytransfer1,energytransfer2},
information
retrieval~\cite{InformationRetrieva1,InformationRetrieva2},
coherence flow~\cite{InformationRetrieva3}  and topological
invariants~\cite{invariant1,invariant2}. A complete
characterization of conservation laws in $\mathcal{PT}$-symmetric
systems has been intensely explored~\cite{RCM,SDM2}. For example,
based on the intertwining
relation~\cite{interwining1,interwining2,SRM},
reference~\cite{RCM} has presented a complete set of conserved
observables for a class of $\mathcal{PT}$-symmetric Hamiltonians
in a single-photon linear optical circuit. Moreover, in the
pseudo-Hermitian representation of quantum mechanics~\cite{SDM1},
reference~\cite{APT} has further implemented a model circuit of a
generic anti-$\mathcal{PT}$-symmetric system. A counterintuitive
energy-difference conserving dynamics has been
observed~\cite{APT}, which is in stark contrast to the standard
Hermitian dynamics keeping the system's total energy constant.
However,  based on biorthogonal quantum mechanics, the
manifestation of Noether's theorem and a complete observation of
conserved quantities in  $\mathcal{PT}$-symmetric systems and
their consequences are still lacking both theoretically and
experimentally.

In this work, we extend  Noether's theorem to a class of
significant $\mathcal{PT}$-symmetric non-Hermitian systems and
introduce a generalized expectation value of a time-independent
operator based on biorthogonal quantum
mechanics~\cite{Biorthogonal1,Biorthogonal2,Biorthogonal3,Biorthogonal4}.
For the $\mathcal{PT}$-symmetric systems considered here,  the
eigenvalues of the $\mathcal{PT}$-symmetric Hamiltonian
$\hat{H}_\mathcal{PT}$  change from purely real numbers to purely
imaginary numbers. Such $\mathcal{PT}$-symmetric systems have been
widely used to investigate the dynamics of non-Hermitian systems
in the presence of balanced gain and
loss~\cite{RCM,Ehrenfest4,photonics1,CriticalPhenomena1,InformationRetrieva1,InformationRetrieva2,InformationRetrieva3}.
Our work shows that the extended Noether's theorem can be used to
deal with conservation law problems about pure states and mixed
states. Remarkably, we find that for an operator $\hat{F}$ without
explicit time dependence, its generalized expectation value is a
constant of motion if $\hat{F}$ presents a standard symmetry in
the $\mathcal{PT}$-symmetry  unbroken regime, or a chiral symmetry
in the $\mathcal{PT}$-symmetry  broken regime. In addition, we
experimentally investigate the extended Noether's theorem in
$\mathcal{PT}$-symmetric single-qubit and two-qubit systems using
an optical setup. Several novel results are found. First, our
experiment demonstrates the existence of the constant of motion.
Second, our experiment reveals that the constant of motion can be
used to judge whether the $\mathcal{PT}$ symmetry of a system is
broken. Last, our experiment reveals the phenomenon of masking
quantum information~\cite{masking1,masking2} in a
$\mathcal{PT}$-symmetric two-qubit system.

\section{Extension of Noether's theorem in $\mathcal{PT}$-symmetric systems}\label{section:II}

To extend Noether's theorem to $\mathcal{PT}$-symmetric systems,
the biorthogonal quantum
mechanics~\cite{Biorthogonal1,Biorthogonal2,Biorthogonal3,Biorthogonal4}
is applied. In biorthogonal quantum mechanics, the inner product
is defined as
\begin{eqnarray}\label{eqm-2}
    (\varphi,\psi)\equiv\langle\widehat{\varphi}|\psi\rangle
    =\sum_{k,l}d^{*}_{k}c_{l}\langle\widehat{{\phi_{k}}}|\phi_{l}\rangle
    =\sum_{k}d^{*}_{k}c_{k},
\end{eqnarray}
where $|\psi\rangle=\Sigma_{l}c_{l}|{\phi_{l}}\rangle$
($|\varphi\rangle=\Sigma_{k}d_{k}|{\phi_{k}}\rangle$) is an
arbitrary pure state with its associated state
$\langle\widehat{\psi}|\equiv\Sigma_{l}c^{*}_{l}\langle\widehat{{\phi_{l}}}|$
($\langle\widehat{\varphi}|\equiv\Sigma_{k}d^{*}_{k}\langle\widehat{{\phi_{k}}}|$),
and $\{\langle\widehat{{\phi_{l(k)}}}|\}$ and
$\{{|\phi_{l(k)}}\rangle\}$ are left and right eigenstates of a
non-Hermitian Hamiltonian~(Appendixes \ref{section:AppendixA}
and \ref{section:AppendixB}).

Here, we use ${\hat{\rho}}$ (${\hat{\rho}}_{b}$) to denote  a
density operator in standard (biorthogonal) quantum mechanics.
Without loss of generality, let us consider the
$\mathcal{PT}$-symmetric system to be in a mixed state
${\hat{\rho}}_{b}(t)=\sum_{n=1}^{N}p_n|\psi_n(t)\rangle\langle\widehat{\psi_n(t)}|$,
$p_n$ is the probability of the system being in a pure state
$|\psi_{n}(t)\rangle$, with
$\langle\widehat{\psi_{n}}(t)|\psi_{n}(t)\rangle=1$. With the
inner product introduced in eq.~(\ref{eqm-2}), a generalized
expectation value $(\hat{F})$ of an operator $\hat{F}$ can be
defined~(see Appendix \ref{section:AppendixC})
\begin{eqnarray}\label{eqm-3}
    (\hat{F})&=&tr[\hat{\rho}_{b}(t){\hat{F}}]
    =\sum_{l}\langle\widehat{{\phi_{l}}}|\hat{\rho}_{b}(t){\hat{F}}|{\phi_{l}}\rangle\cr
    &=&\sum_{n}p_n\langle\widehat{\psi_n(t)}|{\hat{F}}|\psi_n(t)\rangle.
\end{eqnarray}
where $\langle\widehat{\psi_n(t)}|{\hat{F}}|\psi_n(t)\rangle$ is
the generalized expectation value $(\hat{F})$ of the operator
$\hat{F}$ for an arbitrary pure state $|\psi_{n}(t)\rangle$.
Equation~(\ref{eqm-3}) provides a natural generalization of
expectation value of an operator $\hat{F}$ for an arbitrary
quantum state, either a mixed state or a pure state.

As one of the main contributions of this work, we find that the
temporal evolution of the expectation value $(\hat{F})$ of the
operator $\hat{F}$ follows two different forms (see Appendix \ref{section:AppendixC} for the detailed derivation)
\begin{eqnarray}
    \frac{d}{dt}({\hat{F}})=\sum_{n}p_n\left[\frac{1}{i\hbar}([\hat{F},\hat{H}_{\mathcal{PT}}])_{n}+\left({\frac{d\hat{F}}{dt}}\right)_{n}\right],\label{eqm-3a}\\
    \frac{d}{dt}({\hat{F}})=\sum_{n}p_n\left[\frac{1}{i\hbar}(\{\hat{F},\hat{H}_{\mathcal{PT}}\})_{n}+\left({\frac{d\hat{F}}{dt}}\right)_{n}\right],\label{eqm-3b}
\end{eqnarray}
where
$(\cdot)_n=\langle\widehat{\psi_n(t)}|\cdot|\psi_n(t)\rangle$.
Equation~(\ref{eqm-3a}) corresponds to the case when the system
works in the $\mathcal{PT}$-symmetry unbroken regime, while
eq.~(\ref{eqm-3b}) corresponds to the case when the system works
in the $\mathcal{PT}$-symmetry  broken regime. From
eq.~(\ref{eqm-3a}), one can see that the expectation value
$(\hat{F})$ is a constant of motion if the Hamiltonian
$\hat{H}_{\mathcal{PT}}$ and the time-independent operator
$\hat{F}$ satisfy the commutation relation
$[\hat{F},\hat{H}_{\mathcal{PT}}]=0$, i.e., the operator $\hat{F}$
presents a standard symmetry  in the $\mathcal{PT}$-symmetry
unbroken regime~\cite{Symmetry1}. On the other hand,
eq.~(\ref{eqm-3b}) implies that the expectation value $(\hat{F})$
is also a constant of motion if $\hat{H}_{\mathcal{PT}}$ and
$\hat{F}$ satisfy the anti-commutation relation
$\{\hat{F},\hat{H}_{\mathcal{PT}}\}=0$, i.e., $\hat{F}$ presents a
chiral symmetry  in the $\mathcal{PT}$-symmetry  broken
regime~\cite{chiralsymmetry1}.

To understand the above results intuitively, let us consider a
$\mathcal{PT}$-symmetric single-qubit system where the eigenvalues
of the Hamiltonian change from real (in the
$\mathcal{PT}$-symmetry unbroken regime), to purely imaginary (in
the $\mathcal{PT}$-symmetry broken regime). The Hamiltonian for
this system is given by (hereafter, we assume $\hbar=1$)
\begin{equation}\label{eqm-7}
    \hat{H}_{\mathcal{PT}}=s\hat{\sigma}_{x}+i\gamma\hat{\sigma}_{z}=\left(\begin{array}{ll}
        i\gamma & s\\
        s & -i\gamma
    \end{array}\right),
\end{equation}
where $i\gamma\hat{\sigma}_{z}$ is the non-Hermitian part of the
Hamiltonian governing gain and
loss~\cite{Non-Hermitian1,non-Hermitianadd}. The parameter $s>0$
is an energy scale, $a=\gamma/s>0$ is a coefficient representing
the degree of non-Hermiticity, $\hat{\sigma}_{x}$ and
$\hat{\sigma}_{z}$ are the standard Pauli operators. The
eigenvalues  of $\hat{H}_{\mathcal{PT}}$ are given by
$E_{1}=s\sqrt{1-a^{2}}$ and $E_{2}=-s\sqrt{1-a^{2}}$, which  are
real numbers for $0<a<1$ (the $\mathcal{PT}$-symmetry  unbroken
regime), while purely imaginary numbers for $a>1$ (the
$\mathcal{PT}$-symmetry  broken regime). The right eigenvectors of
$\hat{H}_{\mathcal{PT}}$ are
$|\phi_{1}\rangle=f_{1}\times(A_{1}|0\rangle+|1\rangle)$ and
$|\phi_{2}\rangle=f_{2}\times(A_{2}|0\rangle+|1\rangle)$, while
the left eigenvectors of $\hat{H}_{\mathcal{PT}}$ are
$\langle\widehat{\phi}_{1}|=f_{3}^{*}\times(-A_{2}^{*}\langle0|+\langle1|)$
and
$\langle\widehat{\phi}_{2}|=f_{4}^{*}\times(-A_{1}^{*}\langle0|+\langle1|)$~(Appendix \ref{section:AppendixB}). Here, $A_{1}=ia+\sqrt{1-a^{2}}$,
$A_{2}=ia-\sqrt{1-a^{2}}$, and $f_{1},f_{2},f_{3},f_{4}$ satisfy
$f_1\cdot{f}_3^{*}\times(1-A^{*}_{2}A_{1})=f_2\cdot{f}_4^{*}\times(1-A^{*}_{1}A_{2})=1$
to satisfy the biorthogonality and closure relations.

The Hamiltonian~(\ref{eqm-7}) can be considered as a deformed
Pauli operator,
$\hat{H}_{\mathcal{PT}}=E_{1}|{\phi_{1}}\rangle\langle\widehat{{\phi_{1}}}|-E_{1}|{\phi_{2}}\rangle\langle\widehat{{\phi_{2}}}|$,
in view of the biorthogonal partners
$\{|{\phi_{1}}\rangle,|{\phi_{2}}\rangle\}$ and
$\{\langle\widehat{{\phi_{1}}}|,\langle\widehat{{\phi_{2}}}|\}$~(Appendixes \ref{section:AppendixA} and \ref{section:AppendixB}). If
a time-independent operator $\hat{F}$ can be expressed in the form
\begin{equation}
    \hat{F}=c_{1}|{\phi_{1}}\rangle\langle\widehat{{\phi_{1}}}|+c_{2}|{\phi_{2}}\rangle\langle\widehat{{\phi_{2}}}|,\label{eqm-8a}
\end{equation}
where $c_{1}$ and $c_{2}$ are arbitrary nonzero coefficients, one
can easily verify $[\hat{F},\hat{H}_{\mathcal{PT}}]=0$. Thus,
according to eq.~(\ref{eqm-3a}), the expectation value $(\hat{F})$
is a constant of motion in the $\mathcal{PT}$-symmetry unbroken
regime. On the other hand, if a time-independent operator
$\hat{F}$ can be expressed in the form
\begin{equation}
    \hat{F}=\tilde{c}_{1}(|{\phi_{1}}\rangle\langle\widehat{{\phi_{2}}}|-|{\phi_{2}}\rangle\langle\widehat{{\phi_{1}}}|)\label{eqm-8b}
\end{equation}
where $\tilde{c}_{1}$ is an arbitrary nonzero coefficient, one can
obtain $\{\hat{F},\hat{H}_{\mathcal{PT}}\}=0$. In this case,
according to eq.~(\ref{eqm-3b}), the expectation value $(\hat{F})$
is a constant of motion in the $\mathcal{PT}$-symmetry broken
regime.
\begin{figure*}[htbp]
    \center \scalebox{0.80}{\includegraphics{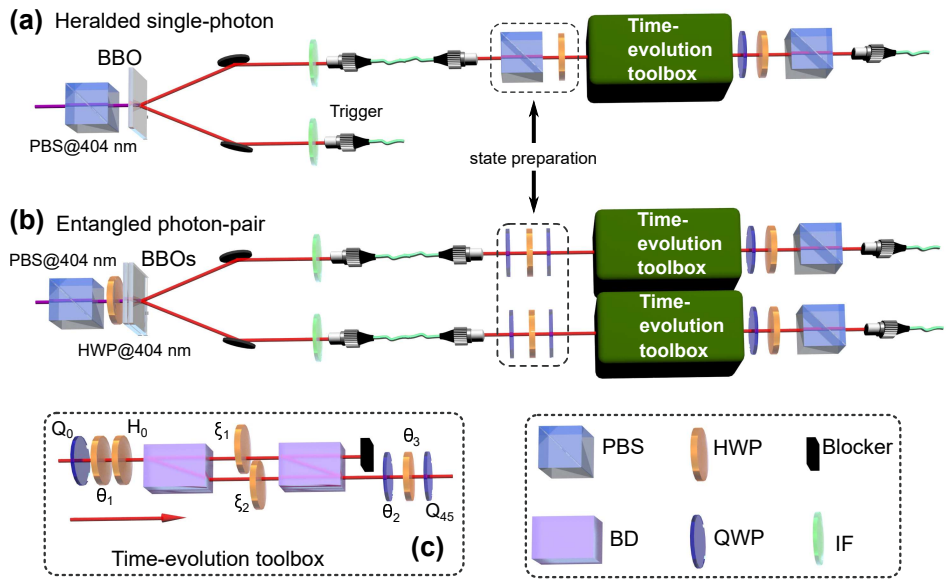}}
    \caption{Experimental setup. (a) Overview of the apparatus for the
        initial state preparation in a single-photon system. For details,
        please refer to the text. (b) Overview of the apparatus for the
        initial state preparation in a two-photon system. For details,
        please refer to the text. (c) Implementation of the time-evolution
        toolbox. Q$_0$ (H$_{0}$) represents a QWP (HWP) with fixed
        orientation $0^{\circ}$, while Q$_{45}$ represents a QWP with
        fixed orientation $\pi/4$. See text for details. PBS: polarization
        beam splitter, HWP: half-wave plate, QWP: quarter-wave plate, BD:
        beam displacer, IF: interference filter, BBO:
        $\beta$-barium-borate crystal.}
    \label{fig1}
\end{figure*}
From an experimental point of view, in
order to keep the expectation value $(\hat{F})$ as a real number, the chosen
operator $\hat{F}$ should be Hermitian in biorthogonal
quantum mechanics~(see Appendix
\ref{section:AppendixD}). Therefore, in the subsequent  discussion, the
coefficients $c_{1}$ and $c_{2}$ in eq.~(\ref{eqm-8a}) are chosen as real
numbers, and the coefficient $\tilde{c}_{1}$ in eq.~(\ref{eqm-8b})
is chosen as a purely imaginary number.

\section{Experimental setup}

\subsection{Single-qubit case}
The apparatus for the initial state preparation in a single-photon
system is illustrated in Figure~\ref{fig1}a, where a single photon
acts as the qubit. A photon pair is generated through a type-I
phase-matched spontaneous parametric down-conversion process. The
idler photon is detected by a single photon detector as a trigger.
The qubit is encoded by the polarization of the heralded single
photon, with $|0\rangle=|H\rangle$ and $|1\rangle=|V\rangle$. The
initial state is prepared by a polarization beam splitter (PBS)
and a half-wave plate (HWP). Then the photon is injected into a
time-evolution toolbox, which outputs the desired time-evolved
state.  In our experiment, the time-evolved state is accessed by
enforcing the time-evolution operator
$\hat{U}_{\mathcal{PT}}(t)$=exp($-i\hat{H}_{\mathcal{PT}}t$) at
any given time on the initial state. Here, the Hamiltonian
$\hat{H}_{\mathcal{PT}}$ is the one given by eq.~(\ref{eqm-7}). As
depicted in Figure~\ref{fig1}c, the time-evolution toolbox
implements the time-evolution operator
${\hat{U}}_{\mathcal{PT}}(t)$ by decomposing it into basic
operations~(see Appendix \ref{section:AppendixE})
\begin{eqnarray}\label{eqm-8}
    \hat{U}_{\mathcal{PT}}(t)&=&\hat{R}_{\textrm{QWP}}(\pi/4)\hat{R}_{\textrm{HWP}}(\theta_3)
    \hat{R}_{\textrm{QWP}}(\theta_2)\hat{L}(\xi_1,~\xi_2) \nonumber\\
    &      &  \hat{R}_{\textrm{HWP}}(0)\hat{R}_{\textrm{HWP}}(\theta_1)\hat{R}_{\textrm{QWP}}(0),
\end{eqnarray}
where the loss-dependent operator
\begin{equation}\label{eqm-9}
    \hat{L}\left(\xi_1,~\xi_2\right)=\left(
    \begin{array}{cc}
        0 & \sin 2\xi_1 \\
        \sin 2\xi_2 & 0 \\
    \end{array}
    \right)
\end{equation}
is realized by a combination of two beam displacers (BDs) and two
HWPs with setting angles $\xi_1$ and $\xi_2$ ($\xi_2$ is fixed
with $\pi/4$ in our experiment). Moreover,
$\hat{R}_{{\textrm{HWP}}}$  and $\hat{R}_{{\textrm{QWP}}}$ are the
rotation operators of the HWP and quarter-wave plate (QWP),
respectively.

The time-evolved states in the $\mathcal{PT}$-symmetric
single-qubit system is given
by~\cite{InformationRetrieva1,rho1,rho2}
\begin{equation}\label{eqm-10}
    \hat{\rho}^{E}(t)=\frac{\hat{U}_{\mathcal{PT}}(t) \hat{\rho}(0)
        \hat{U}_{\mathcal{PT}}^{\dagger}(t)}
    {\textrm{Tr}\left[\hat{U}_{\mathcal{PT}}(t) \hat{\rho}(0)
        \hat{U}_{\mathcal{PT}}^{\dagger}(t)\right]},
\end{equation}
where $\hat{\rho}(0)$ is the initial density matrix and
$\hat{\rho}^{E}(t)$ is the experimental density matrix at any
given time $t$ in standard quantum mechanics. The density matrix
${\rho}^{E}(t)$ can be constructed via quantum state
tomography~\cite{Tomography,Tomography1}. For the single-qubit
system, we project the photon onto 4 bases $\{|H\rangle,
|V\rangle, |R\rangle=(|H\rangle-i|V\rangle)/\sqrt{2},
|D\rangle=(|H\rangle+|V\rangle)/\sqrt{2}\}$. In addition, we note
that the density matrix in biorthogonal quantum mechanics can be
reversely extracted from the density matrix in standard quantum
mechanics $\hat{\rho}^{E}(t)$~(Appendix \ref{section:AppendixF}).
On the other hand, the density matrix ${\hat{\rho}}_{b}(t)$ in
biorthogonal quantum mechanics can be obtained according to the
following relationships~(Appendix \ref{section:AppendixG})
\begin{align}
    {\hat{\rho}}_{b}(t)={\hat{U}_{\mathcal{PT}}(t){\hat{\rho}}_{b}(0)\hat{U}_{\mathcal{PT}}^{'}(t)},\label{eqm-6a}\\
    {\hat{\rho}}_{b}(t)={\hat{U}_{\mathcal{PT}}(t){\hat{\rho}}_{b}(0)\hat{U}_{\mathcal{PT}}(t)},\label{eqm-6b}
\end{align}
where $\hat{U}_{\mathcal{PT}}(t)$=exp($-i\hat{H}_{\mathcal{PT}}t$)
and
$\hat{U}^{'}_{\mathcal{PT}}(t)$=exp($i\hat{H}_{\mathcal{PT}}t$)
are time-evolution operators, and $\hat{\rho}_{b}(0)$ is the
initial density matrix in biorthogonal quantum mechanics.
Equations~(\ref{eqm-6a}) and (\ref{eqm-6b}) correspond to the
cases when the system evolves in the $\mathcal{PT}$-symmetry
unbroken regime and $\mathcal{PT}$-symmetry broken regime,
respectively.

\subsection{Two-qubit case}

The apparatus for the initial state preparation in a two-photon
system is illustrated in Figure~\ref{fig1}b. The entangled states
in the experiment are generated through a type-II phase-matched
spontaneous parametric down-conversion. Then two combinations of
HWPs and QWPs (i.e., the upper and lower parts in the dashed box)
operating on each photon, eliminate the influence caused by the
fibres, therefore preparing the initial state. Then each photon is
injected into a $\mathcal{PT}$-symmetric time evolution toolbox.
The dynamical evolution of quantum states in this case is
similarly given by equation~(\ref{eqm-10}), where the
time-evolution nonunitary operator is now given by
$\hat{U}_{\mathcal{PT}}(t)=\hat{U}_{\mathcal{PT},1}(t)\otimes\hat{U}_{\mathcal{PT},2}(t)$.
Here,
$\hat{U}_{\mathcal{PT},j}(t)=\exp(-i\hat{H}_{\mathcal{PT},j}t)$
($j=1,2$) is the time-evolution nonunitary operator of qubit $j$
in the two-qubit system. Experimentally, we reconstruct the
density matrix $\hat{\rho}^{E}(t)$ at any given time $t$ via
quantum state tomography after each of the two photons passes
through the time-evolution toolbox. Essentially, we project the
two-qubit state onto 16 basis states through a combination of QWP,
HWP and PBS, and then perform a maximum-likelihood estimation of
the density matrix~\cite{Tomography,Tomography1}.

\subsection{ Device parameters}
For the single-qubit case, the photon pair is generated through a
type-I phase-matched spontaneous parametric down-conversion
process by pumping a nonlinear $\beta$-barium-borate (BBO) crystal
with a 404 nm pump laser, where the BBO crystal is 3 mm thick. The
power of the pump laser is 130 mW. The bandwidth of the
interference filter (IF) is 10 nm. This yields a maximum count of
60,000 per second. The quantum state is measured by performing
standard state tomography, i.e.,  projecting the state onto 4
bases $\{|H\rangle, |V\rangle,
|R\rangle=(|H\rangle-i|V\rangle)/\sqrt{2},
|D\rangle=(|H\rangle+|V\rangle)/\sqrt{2}\}$, and the corresponding
angles of QWP-HWP are  $\left(0^{\circ},~0^{\circ}\right)$,
$\left(0^{\circ},~45^{\circ}\right)$,
$\left(45^{\circ},~22.5^{\circ}\right)$,
$\left(0^{\circ},~22.5^{\circ}\right)$, and
$\left(45^{\circ},~0^{\circ}\right)$, respectively.

For the two-qubit case, the entangled states in the experiment are
generated through a type-II phase-matched spontaneous parametric
down-conversion, by pumping two BBO crystals with a 404 nm pump
laser, where each BBO crystal is 0.4 mm thick and the optical axes
are perpendicular to each other. The measurement of the photon
source yields a maximum of 10,000 photon counts over 1.5 s after
the 10 nm IF.  Here, the quantum state is measured by performing
standard state tomography, i.e.,  projecting the state onto 16
bases \{$|HH\rangle$, $|HV\rangle$, $|VV\rangle$, $|VH\rangle$,
$|RH\rangle$, $|RV\rangle$, $|DV\rangle$, $|DH\rangle$,
$|DR\rangle$, $|DD\rangle$, $|RD\rangle$, $|HD\rangle$,
$|VD\rangle$, $|VL\rangle$, $HL\rangle$, $|RL\rangle$\}, where
$|D\rangle = \left(|H\rangle+|V\rangle\right)/\sqrt{2}$,
$|R\rangle = \left(|H\rangle - i |V\rangle\right)/\sqrt{2}$, and
$|L\rangle = \left(|H\rangle+i|V\rangle\right)/\sqrt{2}$.

\section{Experimental and theoretical results}

\begin{figure}[htbp]
\center \scalebox{0.42}{\includegraphics{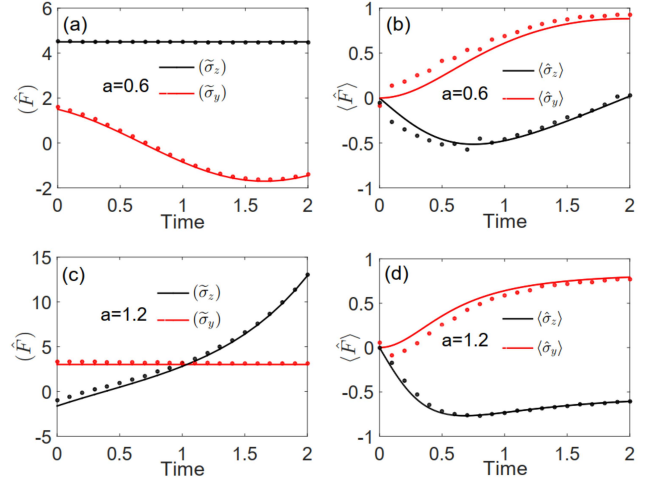}} \caption{ The
temporal evolutions of expectation values $(\hat{F})$ and
$\langle{\hat{F}}\rangle$ in the $\mathcal{PT}$-symmetric
single-photon system. For (a) and (b) the system works in the
$\mathcal{PT}$-symmetry unbroken regime ($a=0.6$), while for (c)
and (d) the system works in the $\mathcal{PT}$-symmetry broken
regime ($a=1.2$). For (a) and (c), the observable operators
$\hat{F}$ are chosen as deformed Pauli operators
$\widetilde{\sigma}_z=|{\phi_{1}}\rangle\langle\widehat{{\phi_{1}}}|-|{\phi_{2}}\rangle\langle\widehat{{\phi_{1}}}|$
and
$\widetilde{\sigma}_y=-i|{\phi_{1}}\rangle\langle\widehat{{\phi_{2}}}|+i|{\phi_{2}}\rangle\langle\widehat{{\phi_{1}}}|$
in biorthogonal quantum mechanics. The expectation value
$(\hat{F})$ is based on
$(\hat{F})={\langle\widehat{{\psi}(t)}|\hat{F}|\psi(t)\rangle}$.
For (b) and (d), the observable operators are chosen as standard
Pauli operators $\hat{\sigma}_z$ and $\hat{\sigma}_y$, the
expectation value $\langle{\hat{F}}\rangle$ is based on
$\langle{\hat{F}}\rangle={\langle{{\psi}(t)}|\hat{F}|\psi(t)\rangle}$.
The initial state is $(|0\rangle+|1\rangle)/\sqrt{2}$, and we have
set $f_{1}=f_{2}=1/\sqrt{2}$ and $s=1$. All curves show the
theoretical results while  dots are the experimental data.}
\label{fig2}
\end{figure}

\subsection{Expectation values of operators in a $\mathcal{PT}$-symmetric single-qubit system}

As two results derived from Noether's theorem,
equations~(\ref{eqm-8a}) and ~(\ref{eqm-8b}) tell us that the
expectation value $(F)$ is a constant of motion if

\begin{equation}\label{eqm-add1}
\hat{F}=\widetilde{\sigma}_z=|{\phi_{1}}\rangle\langle\widehat{{\phi_{1}}}|-|{\phi_{2}}\rangle\langle\widehat{{\phi_{2}}}|,~~
(c_{1}=-c_{2}=1)
\end{equation}
and
\begin{equation}\label{eqm-add2}
\hat{F}=\widetilde{\sigma}_y=-i|{\phi_{1}}\rangle\langle\widehat{{\phi_{2}}}|+i|{\phi_{2}}\rangle\langle\widehat{{\phi_{1}}}|),~~
(\tilde{c}_{1}=-i)
\end{equation}
for the $\mathcal{PT}$-symmetry  unbroken and broken cases,
respectively. We experimentally confirm this prediction in a
$\mathcal{PT}$-symmetric single-qubit system. As shown in
Figure~\ref{fig2}a, in the $\mathcal{PT}$-symmetry  unbroken
regime, $(\widetilde{\sigma}_z)$ is a constant of motion, whereas
$(\widetilde{\sigma}_y)$ changes over time. Interestingly, in
contrast to Figure~\ref{fig2}a, Figure~\ref{fig2}c shows that in
the $\mathcal{PT}$-symmetry  broken regime,
$(\widetilde{\sigma}_y)$ is a constant of motion, while
$(\widetilde{\sigma}_z)$ changes over time. The experimental
results here agree well with the theoretical simulation results.
As a contrast, we also measure the expectation values of
$\hat{\sigma}_z$ and ${\hat{\sigma}}_y$ in standard quantum
mechanics, shown in Figures~\ref{fig2}b and~\ref{fig2}d. One can
see from Figures~\ref{fig2}b and~\ref{fig2}d that both
$\langle{\hat{\sigma}}_z\rangle$ and
$\langle{\hat{\sigma}}_y\rangle$ change over time in the
$\mathcal{PT}$-symmetry  unbroken or broken regime, i.e., one
cannot obtain a constant of motion. Hence, according to the
temporal evolution of expectation values of
$(\widetilde{\sigma}_z)$ and $(\widetilde{\sigma}_y)$, one can
judge whether the system works in the $\mathcal{PT}$-symmetry
unbroken or broken regime.

\begin{figure}[htbp]\center
    \scalebox{0.42}{\includegraphics{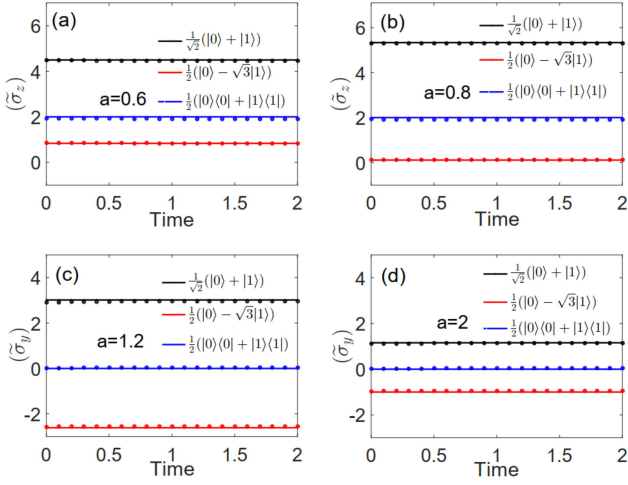}}
    \caption{The temporal evolutions of expectation
        values $(\widetilde{\sigma}_z)$ and  $(\widetilde{\sigma}_y)$ in
        the $\mathcal{PT}$-symmetric single-photon system under different
        initial states and non-Hermiticities. The non-Hermiticities in
        (a), (b), (c) and (d) are chosen as $a=0.6$, $a=0.8$, $a=1.2$ and
        $a=2$, respectively. The initial states are chosen as two pure
        states $\frac{1}{\sqrt{2}}(|0\rangle+|1\rangle)$,
        $\frac{1}{2}(|0\rangle-\sqrt{3}|1\rangle)$  and  a mixed state
        $\frac{1}{2}(|0\rangle\langle0|+|1\rangle\langle1|)$. We have set
        $f_{1}=f_{2}=1/\sqrt{2}$ and $s=1$. All curves show the
        theoretical results while dots are the experimental data.}
        \label{fig3}
\end{figure}

On the other hand, since our experimental apparatus is quite
general and capable of implementing a broad class of nonunitary
operators, we are able to investigate the role of
non-Hermiticities and the effects of initial states on the
temporal evolution of expectation values. It can be clearly seen
from Figures~3a and~3b that with different initial states,
$(\widetilde{\sigma}_z)$ is always a constant in the
$\mathcal{PT}$-symmetry  unbroken regime even though the initial
state is a mixed state. However, the expectation value
$(\widetilde{\sigma}_z)$ is dependent on the initial states.
Comparing Figure~\ref{fig3}a with Figure~\ref{fig3}b, one can see
that the expectation value $(\widetilde{\sigma}_z)$ gradually
increases when the parameter $a$ (representing the degree of
non-Hermiticity) increases. Similarly, Figures~\ref{fig3}c
and~\ref{fig3}d show that in $\mathcal{PT}$-symmetry  broken
regime, $(\widetilde{\sigma}_y)$ is always a {constant} for
different initial states even though the initial state is a mixed
state, and the expectation value $(\widetilde{\sigma}_y)$
gradually decreases when the parameter $a$  increases.

\begin{figure}[htbp]
\center \scalebox{0.42}{\includegraphics{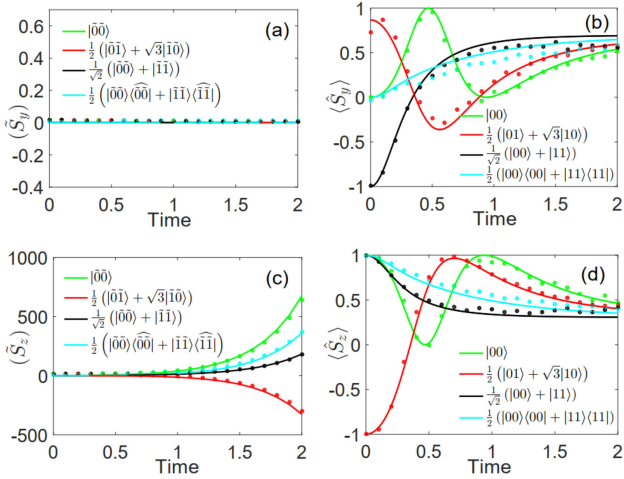}}
\caption{ The
temporal evolutions of expectation values in the
$\mathcal{PT}$-symmetric two-qubit system with different initial
states in the $\mathcal{PT}$ symmetric broken regime ($a=1.2$).
The observable operators in (a) and (b) are chosen as
$\widetilde{S}_y=\widetilde{\sigma}_{y,1}+\widetilde{\sigma}_{y,2}$,
and $\hat{S}_y=\hat{\sigma}_{y,1}+\hat{\sigma}_{y,2}$,
respectively; while the observable operators in (c) and (d) are
chosen as
$\widetilde{S}_z=\widetilde{\sigma}_{z,1}+\widetilde{\sigma}_{z,2}$,
and $\hat{S}_z=\hat{\sigma}_{z,1}+\hat{\sigma}_{z,2}$,
respectively. Here, $\hat{\sigma}_{x,j}$ and $\hat{\sigma}_{z,j}$
($\widetilde{\sigma}_{x,j}$ and $\widetilde{\sigma}_{z,j}$) are
the standard (deformed) Pauli operators for the qubit $j~(j=1,2)$
in standard (biorthogonal) quantum mechanics.
$|\widetilde{0}\rangle\equiv|\phi_{1}\rangle$,
$|\widetilde{1}\rangle\equiv|\phi_{2}\rangle$,
$\langle\widehat{\widetilde{0}}|=\langle\widehat{\phi_{1}}|$,
$\langle\widehat{\widetilde{1}}|=\langle\widehat{\phi_{2}}|$, and
we set $f_{1}=f_{2}=1/\sqrt{2}$ and $s=1$. All curves show the
theoretical results while dots are the experimental data.}
\label{fig4}
\end{figure}

\subsection{Expectation values of operators in a $\mathcal{PT}$-symmetric two-qubit system}

We further study the ${\mathcal{PT}}$ evolution of a two-qubit
system using the optical setup shown in Figure~\ref{fig1}b. The
Hamiltonian of the two-qubit system is described by
$\hat{H}$=$\hat{H}_{\mathcal{PT},1}$+$\hat{H}_{\mathcal{PT},2}$=$s(\hat{S}_{x}$+$ia\hat{S}_{z})$,
with
$\hat{H}_{\mathcal{PT},j}$=$s(\hat{\sigma}_{x,j}$+$ia\hat{\sigma}_{z,j})$,
$\hat{S}_x$=$\hat{\sigma}_{x,1}$+$\hat{\sigma}_{x,2}$, and
$\hat{S}_z$=$\hat{\sigma}_{z,1}$+$\hat{\sigma}_{z,2}$. Here,
$\hat{\sigma}_{x,j}$ and $\hat{\sigma}_{z,j}$ are the standard
Pauli operators for the photonic qubit $j~(j=1,2)$. The parameter
$s$ is still the energy scale. For different initial states, the
temporal evolutions of expectation values in the two-qubit system
are plotted in Figure~\ref{fig4}. The observable operators in
Figure~\ref{fig4}a and Figure~\ref{fig4}c are chosen as
$\widetilde{S}_y$=$\widetilde{\sigma}_{y,1}$+$\widetilde{\sigma}_{y,2}$
and
$\widetilde{S}_z$=$\widetilde{\sigma}_{z,1}$+$\widetilde{\sigma}_{z,2}$,
respectively. Here, $\widetilde{\sigma}_{y,j}$ and
$\widetilde{\sigma}_{z,j}$ are deformed Pauli operators for the
qubit $j~(j=1,2)$ in biorthogonal quantum mechanics.  One can
verify $\{\widetilde{S}_y,\hat{H}\}=0$ and
$[\widetilde{S}_z,\hat{H}]=0$. As expected, Figures~\ref{fig4}a
and~\ref{fig4}c show that $(\widetilde{S}_y)$ remains unchanged,
whereas $(\widetilde{S}_z)$ changes quickly in the
$\mathcal{PT}$-symmetry  broken regime ($a=1.2$). Remarkably, it's
worth noting that the expectation value $(\widetilde{S}_y)$ is
\emph{zero}, which is independent of the initial states. Taking an
information-theoretic perspective on this phenomenon, one can thus
conclude that the information of the initial states is masked when
measuring the expectation value $(\widetilde{S}_y)$, while the
information of the initial states can be disclosed by measuring
the expectation value $(\widetilde{S}_z)$. In addition,
Figures~\ref{fig4}b and~\ref{fig4}d show that both
$\langle{\hat{S}}_z\rangle$ and $\langle{\hat{S}}_y\rangle$ depend
on the initial states and change over time, i.e., the phenomenon
of masking quantum information does not exist in standard quantum
mechanics. Hence, the  masking of quantum information is a unique
phenomenon in biorthogonal quantum mechanics.

\section{CONCLUSION }\label{section:V}

We have extended  Noether's theorem to  a class of significant
$\mathcal{PT}$-symmetric non-Hermitian systems and introduced a
generalized expectation value of a time-independent operator based
on biorthogonal quantum mechanics.  We have demonstrated that in
the $\mathcal{PT}$-symmetry unbroken regime, the generalized
expectation value of a time-independent operator is a constant of
motion, if  the time-independent operator and the non-Hermitian
Hamiltonian satisfy the commutation relation, i.e., the operator
presents a standard symmetry. Moreover, even in the
$\mathcal{PT}$-symmetry broken regime, the expectation value of a
time-independent operator is still a constant of motion provided
the operator and the non-Hermitian Hamiltonian satisfy the
anti-commutation relation, i.e., the operator presents a chiral
symmetry. Furthermore, we have experimentally confirmed our
predictions in $\mathcal{PT}$-symmetric single-qubit and two-qubit
systems by using an optical setup. Our experiment has demonstrated
the existence of the predicted constant of motion. Meanwhile, a
novel phenomenon of masking quantum information is first observed
in a $\mathcal{PT}$-symmetric two-qubit system. The extended
Noether's theorem not only contributes to a full understanding of
the relation between symmetry and conservation law in
$\mathcal{PT}$-symmetric physics, but also has potential
applications in quantum information theory and quantum
communication protocols.

The present work has some elements in common with previous works
on obtaining conserved quantity in non-Hermitian systems,
especially the idea of using pseudo-Hermiticity (equivalently, the
intertwining
relation)~\cite{interwining1,interwining2,SRM,SDM1,SDM2}.
Therefore,  we here address the difference between our work and
previous works.  As shown in
Refs.~\cite{interwining1,interwining2}, every Hamiltonian with a
real spectrum is pseudo-Hermitian, and all the
$\mathcal{PT}$-symmetric non-Hermitian Hamiltonians belong to the
so-called pseudo-Hermitian Hamiltonians. In the pseudo-Hermitian
representation of quantum mechanics, the expectation value
$\langle{\hat{F}}\rangle$ of a time-independent operator
${\hat{F}}$ is a conserved quantity provided the intertwining
relation, $\hat{F}\hat{H}=\hat{H}^\dag{\hat{F}}$, is satisfied. In
principle, a complete set of conserved observables can be obtained
by numerically solving a set of $N^{2}$-dimensional linear
intertwining relation~\cite{SRM,SDM1,SDM2,RCM,SPA}. However, a
common problem, which one may encounter via pseudo-Hermiticity
(intertwining relation), is how to connect the conserved
quantities with the symmetries of dynamics. Compared with previous
works~\cite{interwining1,interwining2,SRM,SDM1,SDM2,RCM,SPA}, the
main difference of our work  is that by introducing a generalized
expectation value of an operator based on biorthogonal quantum
mechanics, we connect two important symmetries with conserved
operators in the $\mathcal{PT}$-symmetry unbroken and broken
regimes, respectively. We remark that the proposed standard
symmetry $\hat{F}\hat{H}=\hat{H}{\hat{F}}$ and the chiral symmetry
$\hat{F}\hat{H}=-\hat{H}{\hat{F}}$ are essentially different from
the intertwining relation $\hat{F}\hat{H}=\hat{H}^\dag{\hat{F}}$,
because of $\hat{H}\neq\hat{H}^\dag$  and
$\hat{H}\neq-\hat{H}^\dag$ in  $\mathcal{PT}$-symmetric systems.

We note that the extended Noether's theorem is always valid for
such  $\mathcal{PT}$-symmetric systems provided the eigenvalues of
$\hat{H}_\mathcal{PT}$ change from purely real numbers to purely
imaginary numbers; or equivalently,  $\hat{H}_{\mathcal{PT}}$
exhibits an exceptional point of the order of the system's
dimension.  As an example, consider a 3-dimensional
$\mathcal{PT}$-symmetric system~\cite{SDM2}, for which the
Hamiltonian reads $H_{\mathcal{PT}}=sJ_{x}+i\gamma J_{z}$, where
$J_{x}$ and $J_{z}$ are the 3-dimensional angular momentum
operators. Such a $\mathcal{PT}$-symmetric Hamiltonian has a
third-order exceptional point at $\gamma=s$ and its spectrum also
changes from  real to purely imaginary~\cite{SDM2}. Then, based on
the extended Noether's theorem, one can quickly find its conserved
quantities in the $\mathcal{PT}$-symmetry unbroken and broken
regimes, respectively.

For any quantum system, whose Hamiltonian can be simplified to the
form in eq.~(\ref{eqm-7}),  the extended Noether's theorem
presented in this work can be implemented straightforwardly. Note
that for the simplified Hamiltonian, arbitrary dressed states can
be chosen as basis states as long as the dressed states satisfy
the biorthogonality and closure relations. This might lead to a
useful step toward realizing fast symmetry discrimination and
conserved quantity acquisition for multi-qubit
$\mathcal{PT}$-symmetric systems. Moreover, in above discussion,
we focus on the case of an operator $\hat{F}$ without explicit
time dependence. However,  the derived
equations~(\ref{eqm-3a},\ref{eqm-3b}) also work well in a general
case i.e., the operator $\hat{F}(t)$ is time-dependent. Then, one
may obtain constant of motion for a time-dependent operator in a
time-dependent $\mathcal{PT}$-symmetric system, which may be
interesting and attractive. Furthermore, in some sense, the
$\mathcal{PT}$-symmetric Hamiltonian in eq.~(\ref{eqm-7}) has
parallels with non-Hermitian topological
phases~\cite{topological1,topological2} and the extended
classification of topological
classes~\cite{chiralsymmetry3,chiralsymmetry4}. The discovery of
the relation between  conserved quantities and non-Hermitian
topological invariants~\cite{invariant1,invariant2} is also
interesting and attractive, which is a fascinating field where
further extension of this work may be explored.

\section*{ACKNOWLEDGEMENT}

This work was supported by the National Natural Science Foundation
of China (NSFC) (Grants Nos.~12264040, 12204311, 11804228,
11865013 and U21A20436), Jiangxi Natural Science Foundation
(20212BAB211018, 20192ACBL20051),  the project of Jiangxi Province
Higher educational Science and Technology Program (Grant
Nos.~GJJ190891, GJJ211735), and Key-Area Research and Development
Program of Guang Dong province (2018B03-0326001). F.N. is
supported in part by Nippon Telegraph and Telephone Corporation
(NTT) Research, the Japan Science and Technology Agency (JST) [via
the Quantum Leap Flagship Program (Q-LEAP), and the Moonshot
R$\&$D Grant Number JPMJMS2061], the Japan Society for the
Promotion of Science (JSPS) [via the Grants-in-Aid for Scientific
Research (KAKENHI) Grant No. JP20H00134], the Army Research Office
(ARO) (Grant No. W911NF-18-1-0358), the Asian Office of Aerospace
Research and Development (AOARD) (via Grant No. FA2386-20-1-4069),
and the Foundational Questions Institute Fund (FQXi) via Grant No.
FQXi-IAF19-06.


\appendix

\section{}

\subsection{Eigenstates of  non-Hermitian Hamiltonians in Biorthogonal quantum mechanics}\label{section:AppendixA}

We first briefly recall some important properties of non-Hermitian
Hamiltonians in biorthogonal quantum
mechanics~\cite{Non-Hermitian1,Non-Hermitian2,Non-Hermitian4,Non-Hermitian5,Biorthogonal1,Biorthogonal2,Biorthogonal3,Biorthogonal4}.
Consider an arbitrary time-independent non-Hermitian Hamiltonian
$\hat{H}$ with $N$ eigenstates $\{|{\phi_{k}}\rangle\}$,
$k=1,2,...,N.$ It satisfies the following eigenvalue equation
\begin{eqnarray}\label{eqa0-1}
    \hat{ H}|{\phi_{k}}\rangle=E_{k}|{\phi_{k}}\rangle.
\end{eqnarray}
As the adjoint operator of $\hat{H}$, the Hamiltonian
$\hat{H}^\dag$  satisfes the following eigenvalue equation
\begin{eqnarray}\label{eqa0-2}
    \hat{H}^\dag|\widehat{{\phi_{k}}}\rangle=E_{k}^{*}|\widehat{{\phi_{k}}}\rangle,
\end{eqnarray}
where $\{|\widehat{{\phi_{k}}}\rangle\}$ are the eigenstates of
$\hat{H}^\dag$ and also the biorthogonal partners of
$\{|{\phi_{k}}\rangle\}$.  The asterisk here means  complex
conjugate. The biorthogonal partners are normalized to satisfy the
biorthogonality
relation~\cite{Biorthogonal1,Biorthogonal2,Biorthogonal3,Biorthogonal4}
\begin{eqnarray}\label{eqa0-3}
    \langle\widehat{{\phi_{k}}}|\phi_{l}\rangle=\delta_{kl},
\end{eqnarray}
and the closure relation
\begin{eqnarray}\label{eqa0-4}
    \sum_{k}|\widehat{{\phi_{k}}}\rangle\langle\phi_{k}|=\sum_{k}|{\phi_{k}}\rangle\langle\widehat{{\phi_{k}}}|=1.
\end{eqnarray}

In this case, if the orthogonality of eigenstates in  standard
quantum mechanics is replaced by the biorthogonality that defines
the relation between the quantum states in the Hilbert space and
its dual space, the resulting quantum theory is called
biorthogonal quantum
mechanics~\cite{Biorthogonal1,Biorthogonal2,Biorthogonal3,Biorthogonal4}. Then,
in  biorthogonal quantum mechanics, the Hamiltonian $\hat{H}$ and
its adjoint Hamiltonian $\hat{H}^\dag$ can be expressed as
\begin{eqnarray}\label{eqa0-5}
    \hat{H}&=&\sum_{k}|{\phi_{k}}\rangle{E_{k}}\langle\widehat{{\phi_{k}}}|,
    \cr
    \hat{H}^\dag&=&\sum_{k}|\widehat{{\phi_{k}}}\rangle{E_{k}^{*}}\langle{\phi_{k}}|.
\end{eqnarray}
For simplicity, $\{\langle\widehat{{\phi_{k}}}|\}$ and
$\{{|\phi_k}\rangle\}$ are called the \textit{left} and
\textit{right} eigenstates of the  Hamiltonian, respectively. In
addition, the overlap distance $\Theta$ between two arbitrary pure
states $|\psi\rangle=\sum_{l}c_{l}|{\phi_{l}}\rangle$ and
$|\varphi\rangle=\sum_{k}d_{k}|{\phi_{k}}\rangle$ can be defined
as~\cite{Biorthogonal1}
\begin{eqnarray}\label{eqc-8}
    \cos^{2}\frac{\Theta}{2}=\frac{\langle\widehat{{\psi}}|{\varphi}\rangle\langle\widehat{{\varphi}}|{\psi}\rangle}
    {\langle\widehat{{\psi}}|{\psi}\rangle\langle\widehat{{\varphi}}|{\varphi}\rangle},
\end{eqnarray}
where
$\langle\widehat{\psi}|=\sum_{l}c^{*}_{l}\langle\widehat{{\phi_{l}}}|$
and
$\langle\widehat{\varphi}|=\sum_{k}d^{*}_{k}\langle\widehat{{\phi_{k}}}|$.
In particular, $\Theta=0$ only if
$|{\psi}\rangle=\pm|{\varphi}\rangle$, whereas $\Theta=\pi$ only
if
$\langle\widehat{{\varphi}}|{\psi}\rangle=\langle\widehat{{\psi}}|{\varphi}\rangle=0$.
{For a two-dimensional Hilbert space, the state $|\psi\rangle$ can
    be expressed in the form
    $|\psi\rangle=\cos\vartheta|{\phi_{1}}\rangle+\sin\vartheta{e^{i\varphi}}|{\phi_{2}}\rangle$,
    with $\langle\widehat{{\psi}}|{\psi}\rangle=1$. The two
    eigenstates $|{\phi_{1}}\rangle$ and $|{\phi_{2}}\rangle$ here can
    be considered as antipodal points on the Bloch sphere. This is
    analogous to the counterpart of a Hermitian system, even though
    $|{\phi_{1}}\rangle$ and $|{\phi_{2}}\rangle$ may  not be
    orthogonal, i.e. $\langle{\phi_{2}}|{\phi_{1}}\rangle\neq0$.}  The usual Bloch sphere description is not adequate at
the exceptional points (EPs). Since at the EPs the intended
antipodal points ($|{\phi_{1}}\rangle$ and $|{\phi_{2}}\rangle$)
completely overlap (i.e.,
$|{\phi_{1}}\rangle$=$|{\phi_{2}}\rangle$), the Bloch sphere will
then become a dot naturally.

\subsection{Eigenstates and eigenvalues of non-Hermitian Hamiltonians in a $\mathcal{PT}$-symmetric single-qubit system}\label{section:AppendixB}

We start with a $\mathcal{PT}$-symmetric non-Hermitian Hamiltonian
in a single-qubit system
\begin{eqnarray}
    \hat{H}_{\mathcal{PT}}=s\hat{\sigma}_{x}+i\gamma\hat{\sigma}_{z}=\left(\begin{array}{ll}
        i\gamma & s\\
        s & -i\gamma
    \end{array}\right),\label{eqc-1}
\end{eqnarray}
where $s\hat{\sigma}_{x}$ is the Hermitian part of the
Hamiltonian, $i\gamma\hat{\sigma}_{z}$ is the non-Hermitian part
of the Hamiltonian governing gain or loss. Moreover, the parameter
$s>0$ is an energy scale, $a=\gamma/s>0$ is a coefficient
representing the degree of non-Hermiticity, $\hat{\sigma}_{x}$ and
$\hat{\sigma}_{z}$ are the standard Pauli operators. The
eigenvalues and eigenvectors of $\hat{H}_{\mathcal{PT}}$ are given
by
\begin{eqnarray}\label{eqc-2}
    E_{1}&=&s\sqrt{1-a^{2}},~~~|\phi_{1}\rangle=f_{1}*(A_{1}|0\rangle+|1\rangle),\cr
    E_{2}&=&-s\sqrt{1-a^{2}},~|\phi_{2}\rangle=f_{2}*(A_{2}|0\rangle+|1\rangle),
\end{eqnarray}
where $A_{1}=ia+\sqrt{1-a^{2}}$, $A_{2}=ia-\sqrt{1-a^{2}}$. Here,
$f_{1}$ and $f_{2}$ are undetermined coefficients. {The
    eigenvalues are  real numbers for $0<a<1$ (the
    $\mathcal{PT}$-symmetry unbroken regime), while imaginary numbers
    for $a>1$ (the $\mathcal{PT}$-symmetry broken regime).} As the
adjoint operator of $\hat{H}_{\mathcal{PT}}$, the eigenvalues and
eigenvectors of $\hat{H}_{\mathcal{PT}}^\dag$ are given by
\begin{eqnarray}\label{eqc-3}
    E^{'}_{1}=s\sqrt{1-a^{2}},~~~|\widehat{\phi}_{1}\rangle=f_{3}*(-A_{2}|0\rangle+|1\rangle),\cr
    E^{'}_{2}=-s\sqrt{1-a^{2}},~|\widehat{\phi}_{2}\rangle=f_{4}*(-A_{1}|0\rangle+|1\rangle),
\end{eqnarray}
where $f_{3}$ and $f_{4}$ are undetermined coefficients. By
substituting eqs.~(\ref{eqc-2}) and (\ref{eqc-3}) into
eq.~(\ref{eqa0-4}), one can find that
\begin{eqnarray}\label{eqc-5}
    f_1\cdot{f}_3^{*}\times(1-A^{*}_{2}A_{1})=1,~~f_2\cdot{f}_4^{*}\times(1-A^{*}_{1}A_{2})=1.
\end{eqnarray}
Theoretically, the coefficients $f_{1},f_{2},f_{3}$ and $f_{4}$
take arbitrary values provided they satisfy the
relation~(\ref{eqc-5}).  However, the values of
$f_{1},f_{2},f_{3}$, and $f_{4}$ may affect the transformation
from the orthogonal space representation to the biorthogonal space
representation.

In the $\mathcal{PT}$-symmetry unbroken regime, the dynamics of
the non-Hermitian single-qubit system will gradually turn into the
dynamics of a Hermitian single-qubit system when  the parameter
$a$ (representing the degree of non-Hermiticity)  tends to zero.
In this case, one can set
\begin{eqnarray}\label{eqc-9}
    \langle{{\phi_{1}}}|{{\phi_{1}}}\rangle=\langle{{\phi_{2}}}|{{\phi_{2}}}\rangle=1,
\end{eqnarray}
so that  $|{\phi_{1}}\rangle$ and $|{\phi_{2}}\rangle$ is in line
with  basis states in the Hermitian single-qubit system. That is,
$|f_{1}|^{2}=1+|A_{2}|^{2},|f_{2}|^{2}=1+|A_{1}|^{2}$. Moreover,
in the $\mathcal{PT}$-symmetry unbroken regime, by setting
$\sin\theta=a$, one can find
\begin{eqnarray}\label{eqc-6}
    A_{1}=\exp(i\theta),~~A_{2}=-\exp(-i\theta), \cr
    \frac{1}{f_1\cdot{f}_3^{*}}=\exp(2i\theta)+1,~~
    \frac{1}{f_2\cdot{f}_4^{*}}=\exp(-2i\theta)+1.
\end{eqnarray}
While, in the $\mathcal{PT}$-symmetry broken regime, by setting
$\sin\theta=1/a$, one has
\begin{eqnarray}\label{eqc-7}
    A_{1}=i\cot\theta^{'},~A_{2}=i\tan\theta^{'},~
    {f_1\cdot{f}_3^{*}}={f_2\cdot{f}_4^{*}}=1/2.
\end{eqnarray}

\subsection{Extended Noether's theorem for a
    $\mathcal{PT}$-symmetric system}\label{section:AppendixC}

Theoretically, there is more than one way to define the inner
product in non-Hermitian systems. In biorthogonal quantum
mechanics, the inner product for a non-Hermitian system is defined
as ~\cite{Biorthogonal1,Biorthogonal2,Biorthogonal3,Biorthogonal4}
\begin{eqnarray}\label{eqa0-6}
    (\varphi,\psi)\equiv\langle\widehat{\varphi}|\psi\rangle
    =\sum_{k,l}d^{*}_{k}c_{l}\langle\widehat{{\phi_{k}}}|\phi_{l}\rangle
    =\sum_{k}d^{*}_{k}c_{k},
\end{eqnarray}
where $|\psi\rangle=\Sigma_{l}c_{l}|{\phi_{l}}\rangle$
($|\varphi\rangle=\Sigma_{k}d_{k}|{\phi_{k}}\rangle$) is an
arbitrary pure state with its associated state
$\langle\widehat{\psi}|\equiv\Sigma_{l}c^{*}_{l}\langle\widehat{{\phi_{l}}}|$
($\langle\widehat{\varphi}|\equiv\Sigma_{k}d^{*}_{k}\langle\widehat{{\phi_{k}}}|$).

Quantum systems are usually characterized by mixed states. Thus,
it is significant to find the  extension of Noether's theorem for
mixed states. For a general $\mathcal{PT}$-symmetric system, its
mixed state  at any given time $t$ can be expressed as a
biorthogonal density operator
\begin{eqnarray}\label{eqb0-4}
    {\hat{\rho}}_{b}(t)=\sum_{n=1}^{N}p_n|\psi_n(t)\rangle\langle\widehat{\psi_n(t)}|.
\end{eqnarray}
Here, $p_n$ is the probability of the system being in the pure
state $|\psi_{n}(t)\rangle$, with
$\langle\widehat{\psi_{n}}(t)|{\psi_{n}(t)}\rangle=1$. Then, for
the case of mixed states, the expectation value $(\hat{F})$ of an
operator $\hat{F}$ is defined as~\cite{Biorthogonal1}
\begin{eqnarray}\label{eqa0-15}
    (\hat{F})&\equiv&tr[\hat{\rho}_{b}(t){\hat{F}}]\cr
    &=&\sum_{m}\langle\widehat{{\phi_{m}}}|\hat{\rho}_{b}(t){\hat{F}}|{\phi_{m}}\rangle\cr
    &=&\sum_{n}\sum_{m}\langle\widehat{{\phi_{m}}}|p_n|\psi_n(t)\rangle\langle\widehat{\psi_n(t)}|{\hat{F}}|{\phi_{m}}\rangle\cr
    &=&\sum_{n}p_n\sum_{m}\langle\widehat{{\phi_{m}}}|\psi_n(t)\rangle\langle\widehat{\psi_n(t)}|{\hat{F}}|{\phi_{m}}\rangle\cr
    &=&\sum_{n}p_n\sum_{m}\langle\widehat{\psi_n(t)}|{\hat{F}}|{\phi_{m}}\rangle\langle\widehat{{\phi_{m}}}|\psi_n(t)\rangle\cr
    &=&\sum_{n}p_n\langle\widehat{\psi_n(t)}|{\hat{F}}|\psi_n(t)\rangle,
\end{eqnarray}
where $\langle\widehat{\psi_n(t)}|{\hat{F}}|\psi_n(t)\rangle$ is
the expectation value $(\hat{F})$ of the operator $\hat{F}$ for an
arbitrary pure state $|\psi_{n}(t)\rangle$. Note that the closure
relation
$\sum_{m}|{\phi_{m}}\rangle\langle\widehat{{\phi_{m}}}|=1$ has
been applied to derive  eq.~(\ref{eqa0-15}).
Equation~(\ref{eqa0-15}) is a natural generalization of the
expectation value of an operator $\hat{F}$ for an arbitrary
quantum state, either a mixed state or a pure state.

Furthermore, consider an arbitrary initial  pure state
$|\psi_{n}(0)\rangle=\Sigma_{k}c_{k}|{\phi_{k}}\rangle$ for a
general $\mathcal{PT}$-symmetric system. According to the
Schr\"{o}dinger equation
\begin{eqnarray}\label{eqa0-9}
    \frac{d|\psi_{n}(t)\rangle}{dt}=\frac{1}{i\hbar}\hat{H}_{\mathcal{PT}}|\psi_{n}(t)\rangle,
\end{eqnarray}
one can obtain the time-evolved state
$|\psi_{n}(t)\rangle$=$\sum_{k}c_{k}e^{-iE_{k}t/\hbar}|{\phi_{k}}\rangle$
at any given time $t$ and its associated state
$\langle\widehat{\psi_{n}(t)}|$=$\sum_{k}c^{*}_{k}e^{iE^{*}_{k}t/\hbar}\langle\widehat{\phi_{k}}|$.

For a general $\mathcal{PT}$-symmetric system, the eigenvalues of
the  $\mathcal{PT}$-symmetric  Hamiltonian
$\hat{H}_{\mathcal{PT}}$ are  real numbers in the
$\mathcal{PT}$-symmetry unbroken regime. Whereas, the eigenvalues
are complex numbers or purely imaginary numbers in the
$\mathcal{PT}$-symmetry broken regime. Thus, in the
$\mathcal{PT}$-symmetry unbroken regime, all the eigenvalues
$\{E_{k}\}$ are real numbers (i.e., $E_{k}$=$E^{*}_{k}$), then
$\langle\widehat{\psi_{n}(t)}|$ satisfies the following
Schr\"{o}dinger equation
\begin{eqnarray}\label{eqa0-10}
    \frac{d\widehat{\langle\psi_{n}(t)|}}{dt}&=&\frac{d{\sum_{k}c^{*}_{k}e^{iE^{*}_{k}t/\hbar}\langle\widehat{\phi_{k}}|}}{dt}\cr
    &=&\sum_{k}\frac{iE^{*}_{k}}{\hbar}c^{*}_{k}e^{iE^{*}_{k}t/\hbar}\langle\widehat{\phi_{k}}|\cr
    &=&\sum_{k}\frac{iE_{k}}{\hbar}c^{*}_{k}e^{iE^{*}_{k}t/\hbar}\langle\widehat{\phi_{k}}|\cr
    &=&\sum_{k}\frac{i}{\hbar}c^{*}_{k}e^{iE^{*}_{k}t/\hbar}[\hat{H}^{\dag}_{\mathcal{PT}}|\widehat{\phi_{k}}\rangle]^\dag\cr
    &=&\frac{1}{-i\hbar}\sum_{k}c^{*}_{k}e^{iE^{*}_{k}t/\hbar}\langle\widehat{\phi_{k}}|\hat{H}_{\mathcal{PT}}\cr
    &=&\frac{1}{-i\hbar}\langle\widehat{\psi_{n}(t)}|\hat{H}_{\mathcal{PT}}.
\end{eqnarray}
Note that the relations
$\hat{H}^\dag|\widehat{{\phi_{k}}}\rangle=E_{k}^{*}|\widehat{{\phi_{k}}}\rangle$
and
$[\hat{H}^{\dag}_{\mathcal{PT}}|\widehat{\phi_{k}}\rangle]^\dag=\langle\widehat{\phi_{k}}|\hat{H}_{\mathcal{PT}}=E_{k}\langle\widehat{\phi_{k}}|$
have been applied.

On the other hand, in the $\mathcal{PT}$-symmetry broken regime,
all the eigenvalues $\{E_{k}\}$ are complex numbers or purely
imaginary numbers. Without loss of generality, consider the
eigenvalue $E_{k}$ with a real part $\textrm{Re}[E_{k}]$ and a
purely imaginary part $\textrm{Im}[E_{k}]$ (i.e.,
$E_{k}$=$\textrm{Re}[E_{k}]$+$i\textrm{Im}[E_{k}]$). Then
$\langle\widehat{\psi_{n}(t)}|$ satisfies the following
Schr\"{o}dinger equation
\begin{eqnarray}\label{eqa0-11}
    &&\frac{d\widehat{\langle\psi_{n}(t)|}}{dt}\cr
    &=&\frac{d{\sum_{k}c^{*}_{k}e^{iE^{*}_{k}t/\hbar}\langle\widehat{\phi_{k}}|}}{dt}\cr
    &=&\sum_{k}\frac{iE^{*}_{k}}{\hbar}c^{*}_{k}e^{iE^{*}_{k}t/\hbar}\langle\widehat{\phi_{k}}|\cr
    &=&\sum_{k}\frac{i(-E_{k}+2\textrm{Re}[E_{k}])}{\hbar}c^{*}_{k}e^{iE^{*}_{k}t/\hbar}\langle\widehat{\phi_{k}}|\cr
    &=&\frac{1}{i\hbar}\sum_{k}c^{*}_{k}e^{iE^{*}_{k}t/\hbar}\langle\widehat{\phi_{k}}|(\hat{H}_{\mathcal{PT}}-2\textrm{Re}[E_{k}])\cr
    &=&\frac{1}{i\hbar}\langle\widehat{\psi_{n}(t)}|\hat{H}_{\mathcal{PT}}
    -\frac{1}{i\hbar}\sum_{k}2\textrm{Re}[E_{k}]c^{*}_{k}e^{iE^{*}_{k}t/\hbar}\langle\widehat{\phi_{k}}|.
\end{eqnarray}
Here we remark that provided $\hat{H}_{\mathcal{PT}}$ exhibits an
exceptional point of the order of the matrix
dimension~\cite{SDM2,Tomography1}, then
$\textrm{Re}[E_{k}]$=$\textrm{Re}[E_{n}]$, $\forall~ k$.
Equation~(\ref{eqa0-11}) can be reduced to
\begin{eqnarray}\label{eqa0-11d}
    \frac{d\widehat{\langle\psi_{n}(t)|}}{dt}
    &=&\frac{1}{i\hbar}\langle\widehat{\psi_{n}(t)}|(\hat{H}_{\mathcal{PT}}-2\textrm{Re}[E_{n}]).
\end{eqnarray}

According to eq.~(\ref{eqa0-15}), the temporal evolution of the
expectation value $(\hat{F})$ can be  expressed as
\begin{eqnarray}\label{eqa0-122}
   && \frac{d}{dt}(\hat{F})\cr
   &=&\sum_{n}p_n\frac{d}{dt}{\langle\widehat{{\psi}_{n}(t)}|\hat{F}|\psi_{n}(t)\rangle}\cr
    &=&\sum_{n}p_n[{\frac{d\langle\widehat{{\psi}_{n}(t)}|}{dt}|\hat{F}|\psi_{n}(t)\rangle}\cr&&
    +{\langle\widehat{{\psi}_{n}(t)}|\hat{F}|\frac{d|\psi_{n}(t)\rangle}{dt}}+{\langle\widehat{{\psi}_{n}(t)}|\frac{d\hat{F}}{dt}|\psi_{n}(t)\rangle}].
\end{eqnarray}
When the eigenvalues of the $\mathcal{PT}$-symmetric Hamiltonian
$\hat{H}_{\mathcal{PT}}$ are real numbers, by substituting
eqs.~(\ref{eqa0-9}) and~(\ref{eqa0-10}) into eq.~(\ref{eqa0-122}),
one can find that the temporal evolution of the expectation value
$(\hat{F})$ reads
\begin{eqnarray}
    \frac{d}{dt}({\hat{F}})&=&\sum_{n}p_n\left[\frac{1}{i\hbar}(\hat{F}\hat{H}_{\mathcal{PT}}-\hat{H}_{\mathcal{PT}}{\hat{F}})_n+\left({\frac{d\hat{F}}{dt}}\right)_n\right]\cr
    &=&\sum_{n}p_n\left[\frac{1}{i\hbar}([\hat{F},\hat{H}_{\mathcal{PT}}])_n
    +\left({\frac{d\hat{F}}{dt}}\right)_n\right]\label{eqm1-3a}
\end{eqnarray}
in the $\mathcal{PT}$-symmetry unbroken regime. Here,
$(\cdot)_n=\langle\widehat{\psi_n(t)}|\cdot|\psi_n(t)\rangle$.

On the other hand, when the eigenvalues of the
$\mathcal{PT}$-symmetric Hamiltonian $\hat{H}_{\mathcal{PT}}$  are
imaginary numbers ($\textrm{Re}[E_{k}]$=0, $\forall~ k$), by
substituting eqs.~(\ref{eqa0-9}) and~(\ref{eqa0-11d}) into
eq.~(\ref{eqa0-122}), one can find that the temporal evolution of
the expectation value $(\hat{F})$ reads
\begin{eqnarray}
    \frac{d}{dt}({\hat{F}})&=&\sum_{n}p_n\left[\frac{1}{i\hbar}(\hat{F}\hat{H}_{\mathcal{PT}}+\hat{H}_{\mathcal{PT}}{\hat{F}})_n+\left({\frac{d\hat{F}}{dt}}\right)_n\right]\cr
    &=&\sum_{n}p_n\left[\frac{1}{i\hbar}(\{\hat{F},\hat{H}_{\mathcal{PT}}\})_n+\left({\frac{d\hat{F}}{dt}}\right)_n\right]\label{eqm1-3b}
\end{eqnarray}
in the $\mathcal{PT}$-symmetry broken regime. One can see that
eq.~(\ref{eqm1-3a}) is eq.~(4) in the main text, while
eq.~(\ref{eqm1-3b}) is eq.~(5) in the main text.

However, if the eigenvalues of the $\mathcal{PT}$-symmetric
Hamiltonian $\hat{H}_{\mathcal{PT}}$ are not purely imaginary
numbers (i.e., $\textrm{Re}[E_{n}]\neq0$),  then by substituting
eqs.~(\ref{eqa0-9}) and~(\ref{eqa0-11d}) into
eq.~(\ref{eqa0-122}), one can find that the temporal evolution of
the expectation value $(\hat{F})$ reads
\begin{eqnarray}
\frac{d}{dt}({\hat{F}})&=&\sum_{n}p_n[\frac{1}{i\hbar}(\hat{F}\hat{H}_{\mathcal{PT}}+\hat{H}_{\mathcal{PT}}{\hat{F}}
\cr&&-2\textrm{Re}[E_{n}]{\hat{F}})_n+\left({\frac{d\hat{F}}{dt}}\right)_n]\label{eqm-3c}
\end{eqnarray}
in the $\mathcal{PT}$-symmetry broken regime. In this case, even
if $\hat{H}_{\mathcal{PT}}$ and $\hat{F}$ satisfy the
anti-commutation relation $\{\hat{H}_{\mathcal{PT}},\hat{F}\}=0$,
the expectation value $(\hat{F})$ is not a constant of motion.

Therefore, in order to obtain a conserved expectation value
$(\hat{F})$ and connect the chiral symmetry with the conserved
operator in the $\mathcal{PT}$-symmetry broken regime, for the
$\mathcal{PT}$-symmetric systems considered in this work, the
eigenvalues of $\hat{H}_\mathcal{PT}$ should change from real
numbers to purely imaginary numbers. We  note that such
${\mathcal{PT}}$-symmetric systems have been widely used to
investigate the dynamics of non-Hermitian systems in the presence
of balanced gain and
loss~\cite{RCM,Ehrenfest4,photonics1,CriticalPhenomena1,InformationRetrieva1,InformationRetrieva2,InformationRetrieva3,rho1,rho2}.
In these cases, the extended Noether's theorem presented in our
work applies well.

\subsection{Conditions for obtaining real expectation values in a $\mathcal{PT}$-symmetric system}\label{section:AppendixD}

From an experimental point of view, it is preferable to keep
expectation values as real numbers. In the following, we will
briefly explore some  conditions for obtaining real expectation
values in a $\mathcal{PT}$-symmetric system.

In standard quantum mechanics, consider a  $N$-dimensional Hilbert
space
\begin{eqnarray}\label{eqa00-1}
\mathscr{H}_S=\textrm{Span}\{|\phi^{'}_{1}\rangle,|\phi^{'}_{2}\rangle,...,|\phi^{'}_{N}\rangle\},
\end{eqnarray}
where the basis state $|\phi^{'}_{k}\rangle$, ($k=1,2,...,N$),
satisfies the orthogonality relation
 \begin{eqnarray}\label{eqa00-2}
 \langle{{\phi^{'}_{k}}}|\phi^{'}_{l}\rangle=\delta_{kl},
 \end{eqnarray}
and the closure relation
 \begin{eqnarray}\label{eqa00-3}
\sum_{k=1}^{N}|{\phi^{'}_{k}}\rangle\langle{{\phi^{'}_{k}}}|=1.
 \end{eqnarray}
Note that the basis state $|\phi^{'}_{k}\rangle$ here is not the
eigenstate of the $\mathcal{PT}$-symmetric Hamiltonian.

A time-independent operator $\hat{F}$ can be expressed by a
density operator
 \begin{eqnarray}\label{eqa00-4}
 \hat{F}=\sum_{k,l}{F}_{kl}|\phi^{'}_{k}\rangle\langle{\phi^{'}_{l}}|,
 \end{eqnarray}
where ${F}_{kl}=\langle{\phi^{'}_{k}}|\hat{F}|\phi^{'}_{l}\rangle$
is the density matrix element of the  operator  $\hat{F}$. Suppose
that the time-evolved state of the $\mathcal{PT}$-symmetric system
reads $|\psi_{n}(t)\rangle=\sum_{k}D_{k}(t)|{\phi^{'}_{k}}\rangle$
at any given time $t$ and its associated state is
$\langle{\psi_{n}(t)}|=\sum_{k}D^{*}_{k}(t)\langle{\phi^{'}_{k}}|$.
Here, $D_{k}(t)$ is a time-dependent and undetermined coefficient.
Then, the standard expectation value $\langle{\hat{F}}\rangle$ for
the pure state $|\psi_{n}(t)\rangle$ reads
 \begin{eqnarray}\label{eqa00-5}
 \langle{\hat{F}}\rangle&=&\langle{\psi_{n}(t)}|\hat{F}|\psi_{n}(t)\rangle\cr
 &=&\sum_{i}D^{*}_{i}(t)\langle{\phi^{'}_{i}}|\sum_{k,l}{F}_{kl}|\phi^{'}_{k}\rangle\langle{\phi^{'}_{l}}|\sum_{j}D_{j}(t)|{\phi^{'}_{j}}\rangle\cr
 &=&\sum_{k,l}D^{*}_{k}(t){F}_{kl}D_{l}(t)\cr
 &=&\sum_{k}|D_{k}(t)|^{2}{F}_{kk}+\sum_{k\neq{l}}D^{*}_{k}(t)D_{l}(t){F}_{kl}.
 \end{eqnarray}

If the time-independent operator $\hat{F}$ is Hermitian in the
$N$-dimensional Hilbert space
 \begin{eqnarray}\label{eqa00-6}
 \hat{F}=\sum_{k,l}{F}_{kl}|\phi^{'}_{k}\rangle\langle{\phi^{'}_{l}}|= \hat{F}^{\dag}=\sum_{k,l}{F}^{*}_{lk}|\phi^{'}_{k}\rangle\langle{\phi^{'}_{l}}|,
 \end{eqnarray}
one can obtain that ${F}_{kk}$ should be a real number and
${F}_{kl}={F}^{*}_{lk}$ $(k\neq{l})$. In this case, the standard
expectation value $\langle{\hat{F}}\rangle$  [see
eq.~(\ref{eqa00-5})] must be a real number, because
$|D_{k}(t)|^{2}{F}_{kk}$ is real and
 \begin{eqnarray}\label{eqa00-7}
&&\sum_{k\neq{l}}D^{*}_{k}(t)D_{l}(t){F}_{kl}\cr
&=&\sum_{k\neq{l},k<{l}}\left[D^{*}_{k}(t)D_{l}(t){F}_{kl}+D^{*}_{l}(t)D_{k}(t){F}_{lk}\right]\cr
&=&\sum_{k\neq{l},k<{l}}\left[D^{*}_{k}(t)D_{l}(t){F}_{kl}+(D^{*}_{k}(t)D_{l}(t){F}_{kl})^{*}\right]\cr
&=&\sum_{k\neq{l},k<{l}}2\textrm{Re}[D^{*}_{k}(t)D_{l}(t){F}_{kl}],
 \end{eqnarray}
where the relation ${F}_{lk}={F}^{*}_{kl}$ $(k\neq{l})$ has been
applied. Thus, the condition for obtaining a real standard
expectation value $\langle{\hat{F}}\rangle$ in a
$\mathcal{PT}$-symmetric system is that the chosen operator
$\hat{F}$ is Hermitian in standard quantum mechanics.

In a similar way, one can prove that the condition for obtaining a
real biorthogonal expectation value $(\hat{F})$ in a
$\mathcal{PT}$-symmetric system is that the chosen operator
$\hat{F}$ is Hermitian in biorthogonal quantum mechanics. Here, we
note that in biorthogonal quantum mechanics, the biorthogonality
relation and the closure relation [see eqs.~(\ref{eqa0-3}) and
~(\ref{eqa0-4})] are applied. A time-independent operator
$\hat{F}$ can be expressed by a biorthogonal density operator
\begin{eqnarray}\label{eqa00-8}
\hat{F}=\sum_{k,l}{F}_{kl}|{\phi_{k}}\rangle\langle{\widehat{\phi_{l}}}|,
\end{eqnarray}
where
${F}_{kl}=\langle{\phi_{k}}|\hat{F}|\widehat{\phi_{l}}\rangle$ is
the biorthogonal density matrix element of the  operator
$\hat{F}$. Moreover, according to eq.~(\ref{eqa0-15}), the
biorthogonal expectation value $(\hat{F})$  reads
\begin{eqnarray}\label{eqa00-9}
(\hat{F})&=&\sum_{n}p_n\langle\widehat{\psi_n(t)}|{\hat{F}}|\psi_n(t)\rangle\cr
&=&\sum_{n}p_n\sum_{i}C^{*}_{i}(t)\langle{\widehat{\phi_{i}}}|\sum_{k,l}{F}_{kl}|{\phi_{k}}\rangle\langle{\widehat{\phi_{l}}}|\sum_{j}C_{j}(t)|{\phi_{j}}\rangle\cr
&=&\sum_{n}p_n\sum_{k,l}C^{*}_{k}(t){F}_{kl}C_{l}(t)\cr
&=&\sum_{n}p_n[\sum_{k}|C_{k}(t)|^{2}{F}_{kk}+\sum_{k\neq{l}}C^{*}_{k}(t)C_{l}(t){F}_{kl}].
\end{eqnarray}
where the time-evolved state
$|\psi_{n}(t)\rangle=\sum_{k}C_{k}(t)|{\phi_{k}}\rangle$  and its
associated state
$\langle\widehat{\psi_{n}(t)}|=\sum_{k}C^{*}_{k}(t)\langle\widehat{\phi_{k}}|$
with $C_{k}(t)=c_{k}e^{-iE_{k}t/\hbar}$ can be obtained from
eq.~(\ref{eqa0-9}).

If the time-independent operator $\hat{F}$ is Hermitian in the
biorthogonal Hilbert space
\begin{eqnarray}\label{eqa00-10}
\hat{F}=\sum_{k,l}{F}_{kl}|{\phi_{k}}\rangle\langle{\widehat{\phi_{l}}}|=
\hat{F}^{\dag}=\sum_{k,l}{F}^{*}_{lk}|{\phi_{k}}\rangle\langle{\widehat{\phi_{l}}}|,
\end{eqnarray}
one can obtain that ${F}_{kk}$ is a real number and also
${F}_{kl}={F}^{*}_{lk}$ $(k\neq{l})$. Then, the  biorthogonal
expectation value $(\hat{F})$  [see eq.~(\ref{eqa00-9})] must be a
real number, because $p_n$ and $|C_{k}(t)|^{2}{F}_{kk}$ are real
and
\begin{eqnarray}\label{eqa00-11}
&&\sum_{k\neq{l}}C^{*}_{k}(t)C_{l}(t){F}_{kl}\cr
&=&\sum_{k\neq{l},k<{l}}\left[C^{*}_{k}(t)C_{l}(t){F}_{kl}+C^{*}_{l}(t)C_{k}(t){F}_{lk}\right]\cr
&=&\sum_{k\neq{l},k<{l}}\left[C^{*}_{k}(t)C_{l}(t){F}_{kl}+(C^{*}_{k}(t)C_{l}(t){F}_{kl})^{*}\right]\cr
&=&\sum_{k\neq{l},k<{l}}2\textrm{Re}[C^{*}_{k}(t)C_{l}(t){F}_{kl}],
\end{eqnarray}
where the relation ${F}_{lk}={F}^{*}_{kl}$ $(k\neq{l})$ has been
applied. That is, the condition for obtaining a real biorthogonal
expectation value $(\hat{F})$ in a $\mathcal{PT}$-symmetric system
is that the chosen operator $\hat{F}$ is Hermitian in biorthogonal
quantum mechanics.

Therefore, in the main text, in order to ensure that the chosen
operators $\hat{F}$ in eqs.~(7,8) are Hermitian in biorthogonal
quantum mechanics, the coefficients $c_{1}$ and $c_{2}$ in eq.~(7)
are real numbers, and the coefficient $\tilde{c}_{1}$ in eq.~(8)
is a purely imaginary number. In addition, when we experimentally
investigate the ``biorthogonal'' expectation value $(\hat{F})$,
the two deformed Pauli operators $\widetilde{\sigma}_z$ and
$\widetilde{\sigma}_y$ (which are Hermitian in biorthogonal
quantum mechanics) are applied. When we experimentally investigate
the standard expectation value $\langle{\hat{F}}\rangle$, the two
standard Pauli operators $\hat{\sigma}_z$ and $\hat{\sigma}_y$
(which are  Hermitian in standard quantum mechanics) are chosen.

\subsection{Decomposition of the nonunitary time-evolution operator}\label{section:AppendixE}

The dynamic evolution of a $\mathcal{PT}$-symmetric single-qubit
system is characterized by the nonunitary time-evolution operator
$U_{\mathcal{PT}}=\exp(-i \hat{H}_{\mathcal{PT}})$, with the
$\mathcal{PT}$-symmetric Hamiltonian $\hat{H}_{\mathcal{PT}} =
s(\hat{\sigma}_{x} + i a \hat{\sigma}_{ z})$. Without loss of
generality, we set $s=1$. In our experiment, we implement the
nonunitary time-evolution operator $U_{\mathcal{PT}}$ by
decomposing it into basic operators.

Let us start with :
\begin{eqnarray}
    \hat{U}_{\mathcal{PT}}(t) & = & \exp(-i\hat{H}_{\mathcal{PT}}t)\nonumber\\
    & = & \exp\left[-i( \sigma_{x}+i a \sigma_{z})t\right]\nonumber\\
    & = &\exp\left[ \left(\begin{array}{cc}
        a & -i\\
        -i    & -a
    \end{array}
    \right)t\right]\nonumber\\
    & = &\left(\begin{array}{cc}
        A+B & -i C\\
        -i  C    & A-B
    \end{array}
    \right).\label{Uapt}
\end{eqnarray}
Here $A$, $B$ and $C$ are given by:

(i) for $0<a<1$,
\begin{equation}
    A=\cos\left(\omega  t\right),~~B=\frac{a}{\omega_{}}\sin\left(\omega t\right),~~C=\frac{1}{\omega}\sin\left(\omega t\right),\label{Eq:APTABC}
\end{equation}
where $\omega=\sqrt{1 - a^2} >0$.

(ii) for $a\geq1$,
\begin{equation}
    A=\cosh\left(\omega t\right),~~B=\frac{a}{\omega}\sinh\left(\omega
    t\right),~~C=\frac{1}{\omega}\sinh\left(\omega
    t\right),\label{Eq:APTABC2}
\end{equation}
where $\omega=\sqrt{a^2-1} \geq0$.

We set the parameters
\begin{eqnarray}
    A   & = & \frac{1}{2}\left(\lambda_2+\lambda_1\right)\sin(-2\theta_1+\theta_2-\pi/4)\label{Eq35},\\
    B   & = &
    \frac{1}{2}\left(\lambda_2-\lambda_1\right)\sin(2\theta_1+\theta_2-\pi/4)\label{Eq36},\\
C&=&-[\lambda_2\sin2\theta_1\cos(\theta_2+\pi/4)\cr&&
    +\lambda_1\cos 2\theta_1\sin\left(\theta_2+\pi/4\right)],\label{Eq37}
\end{eqnarray}
\begin{eqnarray}
    \theta_2&=&\left(2k_1+\frac{3}{4}\right)\pi-2\theta_1,\label{Eq38}\\
    \theta_3&=&\left(\frac{k_2}{2}+\frac{1}{8}\right)\pi-\theta_1,
    \label{Eq39}
\end{eqnarray}
where $k_1$ and $k_2$ are integers. Base on
eqs.~(\ref{Eq35}-\ref{Eq39}), the parameters $\lambda_1$,
$\lambda_2$, $\theta_1$, $\theta_2$ and $\theta_3$ can be
determined with given $A$, $B$ and $C$. The matrix (\ref{Uapt})
can thus be decomposed as follows:
\begin{eqnarray}
    {\hat{U}}_{\mathcal{PT}}(t)
    &=&\left(
    \begin{array}{cc}
        U_{11} &  U_{12}\\
        U_{21} & U_{22} \\
    \end{array}
    \right)\left(
    \begin{array}{cc}
        0 &  \lambda_{1} \\
        \lambda_{2} & 0 \\
    \end{array}
    \right) \left(
    \begin{array}{cc}
        1 & 0\\
        0 & -1\\
    \end{array}
    \right) \cr&& \left(
    \begin{array}{cc}
        \cos 2 \theta_1 &  \sin 2\theta_1\\
        \sin 2 \theta_1 & -\cos 2\theta_1\\
    \end{array}
    \right) \left(
    \begin{array}{cc}
        1 & 0\\
        0 & i\\
    \end{array}
    \right)\label{Uaptsp1},
\end{eqnarray}
where
\begin{eqnarray}
    U_{11} & =& \frac{i}{\sqrt{2}}e^{-i \pi/4} \left(\sin\theta_2 +\cos\theta_2\right) e^{i(\theta_2-2\theta_3)},\\
    U_{12} & =& \frac{i}{\sqrt{2}}e^{-i \pi/4} \left(\sin\theta_2 -\cos\theta_2\right) e^{i(\theta_2-2\theta_3)},\\
    U_{21} & =& \frac{1}{\sqrt{2}}e^{-i \pi/4} \left(\sin\theta_2 -\cos\theta_2\right) e^{-i(\theta_2-2\theta_3)},\\
    U_{22} & =& \frac{1}{\sqrt{2}}e^{-i \pi/4}
    \left(\sin\theta_2+\cos\theta_2\right) e^{-i(\theta_2-2\theta_3)}.
\end{eqnarray}

A half-wave plate (HWP) and a quarter-wave plate (QWP) realize
rotation operations, which are described by the following
operators:
\begin{eqnarray}
    \hat{R}_{ \textrm{QWP}}(\alpha)=\left(\begin{array}{cc}
        \cos^2 \alpha + i\sin^2 \alpha &   ({\sin2\alpha\cos\alpha})/{2}\\
        ({\sin 2\alpha\cos\alpha})/{2} & \sin^2 \alpha + i\cos^2 \alpha\\
    \end{array}\right),\label{RQWP}
\end{eqnarray}
\begin{eqnarray}
    \hat{R}_{ \textrm{HWP}}\left(\beta\right) & = & \left(\begin{array}{cc}
        \cos 2 \beta & \sin 2 \beta\\
        \sin 2 \beta & -\cos 2 \beta\\
    \end{array} \right),\label{RHWP}
\end{eqnarray}
where $\alpha$ and $\beta$ are tunable setting angles. Based on
eq.~(\ref{RQWP}) and eq.~(\ref{RHWP}), we have:
\begin{eqnarray}\label{HWPU1}
    &&\hat{R}_{\textrm{QWP}}(45^{\circ})\hat{R}_{\textrm{HWP}}(\theta_{3})\hat{R}_{\textrm{QWP}}(\theta_2)\cr
    &=&\left(\begin{array}{cc}
        1+i & 1-i\\
        1-i & 1+i\\
    \end{array}\right)
    \left(\begin{array}{cc}
        \cos 2 \theta_3 & \sin 2 \theta_3\\
        \sin 2 \theta_3 & -\cos 2 \theta_3\\
    \end{array}\right)\cr&&\times
    \left(\begin{array}{cc}
        \cos^2 \theta_2 + i\sin^2 \theta_2 & \sin \theta_2 \cdot \cos \theta_2 \left(1-i\right) \\
        \sin \theta_2 \cdot \cos \theta_2 \left(1-i\right) & \sin^2 \theta_2 + i\cos^2 \theta_2\\
    \end{array}\right)\cr
    &=&\left(\begin{array}{cc}
        U_{11}&U_{12}\\
        U_{21} & U_{22}\\
    \end{array}\right),
\end{eqnarray}
\begin{eqnarray}
    {\hat{R}_{ \textrm{HWP}}\left(0^{\circ}\right)  = \left(\begin{array}{cc}
            1 &  0 \\
            0 &   -1 \\
        \end{array}\right),  }
\end{eqnarray}
\begin{eqnarray}
    {\hat{R}_{ \textrm{QWP}}\left(0^{\circ}\right)  = \left(\begin{array}{cc}
            1 &  0 \\
            0 &   i \\
        \end{array}\right),  }
\end{eqnarray}
\begin{eqnarray}
    { \hat{R}_{ \textrm{HWP}}\left(\theta_3\right)= \left(\begin{array}{cc}
            \cos 2 \theta_3 & \sin 2 \theta_3\\
            \sin 2 \theta_3 & -\cos 2 \theta_3\\
        \end{array}\right).}\label{RHWP3}
\end{eqnarray}
After inserting eqs.~(\ref{HWPU1}-\ref{RHWP3}) into
eq.~(\ref{Uaptsp1}), we obtain:
\begin{eqnarray}
    \hat{U}_{\mathcal{PT}}&=&\hat{R}_{\textrm{QWP}}(\pi/4)\hat{R}_{\textrm{HWP}}(\theta_3)
    \hat{R}_{\textrm{QWP}}(\theta_2)\hat{M}(\xi_1,~\xi_2)\cr&&
    \hat{R}_{\textrm{HWP}}(0)\hat{R}_{\textrm{HWP}}(\theta_1)\hat{R}_{\textrm{QWP}}(0),\label{uasp}
\end{eqnarray}
with
\begin{eqnarray}
    {\hat{M}} & = & \left(\begin{array}{cc} 0 &  \lambda_{1} \\
        \lambda_{2} & 0  \end{array}\right).\label{eq:M}
\end{eqnarray}

The matrix ${\hat{M}}$ can be expressed as:
\begin{equation}
    {\hat{M}}=c\left(
    \begin{array}{cc}
        0 & \sin 2\xi_{1} \\
        \sin 2\xi_{2} & 0 \\
    \end{array}
    \right),
\end{equation}
where
$c={\lambda_{1}}/{\sin{2\xi_{1}}}={\lambda_{2}}/{\sin{2\xi_{2}}}$
is a trivial constant. For simplicity, we define:
\begin{equation}
    \hat{L}\left(\xi_{1}, \xi_{2}\right)=\left(
    \begin{array}{cc}0 &  \sin 2\xi_{1} \\
        \sin 2\xi_{2} & 0 \end{array}\right).\label{loss}
\end{equation}
Thus, we have ${\hat{M}}=c\hat{L}$. Note that the functions of
both operators $\hat{L}$ and $c\hat{L}$ are identical. This is
because the states $\hat{L}|\psi\rangle $ and $c\hat{L}
|\psi\rangle$, obtained by enforcing the two operators $\hat{L}$
and $c\hat{L}$ on an arbitrary  state $|\psi\rangle$, are the same
according to the principles of quantum mechanics. Therefore, we
can replace ${\hat{M}}$ in eq.~(\ref{uasp}) by the operator
$\hat{L}$. In this sense, we have from eq.~(\ref{uasp}):
\begin{eqnarray}
    \hat{U}_{\mathcal{PT}}&=&\hat{R}_{\textrm{QWP}}(\pi/4)\hat{R}_{\textrm{HWP}}(\theta_3)
    \hat{R}_{\textrm{QWP}}(\theta_2)\hat{L}(\xi_1,~\xi_2)\cr&&
    \hat{R}_{\textrm{HWP}}(0)\hat{R}_{\textrm{HWP}}(\theta_1)\hat{R}_{\textrm{QWP}}(0),\label{uasp1}
\end{eqnarray}
which is exactly the same as the decomposition of the nonunitary
time-evolution operator $\hat{U}_{\mathcal{PT}}$, described by
eq.~(9) in the main text.

\subsection{Reverse extraction of quantum information in biorthogonal quantum mechanics}\label{section:AppendixF}

Although the mathematical expressions of a given quantum state are
different in standard quantum mechanics and biorthogonal quantum
mechanics, the physical meaning of the given quantum state must be
the same. Based on this idea, for a given quantum state, one can
obtain a one-to-one corresponding relation between the density
matrix in standard quantum mechanics and the density matrix in
biorthogonal quantum mechanics.

For instance, in the orthogonal representation for standard
quantum mechanics,  a quantum state at any given time $t$ can be
given by a density operator
\begin{eqnarray}\label{eq58}
    \hat{\rho}(t)&=&\sum_{n,m}{\rho}_{nm}(t)|{n}\rangle\langle{m}|\cr
    &=&\sum_{n}\lambda_n|\varphi_n(t)\rangle\langle\varphi_n(t)|.
\end{eqnarray}
Note that $\{{\rho}_{nm}(t)\}$ are the density matrix elements of
the density operator $\hat{\rho}(t)$  at any given time $t$ in
standard quantum mechanics,  which can be experimentally obtained
via quantum state tomography. Then, based on the obtained density
matrix elements $\{{\rho}_{nm}(t)\}$, one can  calculate the
eigenvalues $\{\lambda_n\}$ and eigenstates
$\{|\varphi_{n}(t)\rangle\}$ of the density operator
$\hat{\rho}(t)$.

On the other hand, according to biorthogonal quantum mechanics,
the density operator $\hat{\rho}_{b}(t)$ of a quantum state at any
given time $t$ in biorthogonal representation can be expressed as
\begin{eqnarray}\label{eq59}
    \hat{\rho}_{b}(t)&=&\sum_{n}\lambda_n|\varphi_n(t)\rangle\langle\widehat{\varphi_n(t)}|\cr
    &=&\sum_{n,m}\widetilde{\rho}_{nm}(t)|{\phi_n}\rangle\langle{\widehat{{\phi_{m}}}}|,
\end{eqnarray}
where
$\widetilde{\rho}_{nm}(t)$=$\langle{\widehat{{\phi_{m}}}}|\hat{\rho}_{b}(t)|{\phi_n}\rangle$
carries the key quantum information of a  quantum state in
biorthogonal quantum mechanics. Note that the eigenvalues
$\{\lambda_n\}$ and the eigenstates $\{|\varphi_{n}(t)\rangle\}$
can be obtained from eq.~(\ref{eq58}), while
$\{\langle\widehat{{\phi_{m}}}|\}$ and $\{{|\phi_n}\rangle\}$ are
the \textit{left} and \textit{right} eigenstates of the
non-Hermitian Hamiltonian of the system, and they can be obtained
from eqs.~(\ref{eqa0-1}) and ~(\ref{eqa0-2}). In this way, we can
reversely extract the exact information $\widetilde{\rho}_{nm}(t)$
(in biorthogonal quantum mechanics) of a given quantum state from
its density operator in standard quantum mechanics.

\subsection{Dynamical evolution of a class of $\mathcal{PT}$-symmetric
    systems in biorthogonal quantum
    mechanics}\label{section:AppendixG}

Note that the dynamical evolution of a class of
$\mathcal{PT}$-symmetric systems in biorthogonal quantum mechanics
is quite different from that in standard quantum mechanics. In
biorthogonal quantum mechanics, a mixed state
${\hat{\rho}}_{b}(t)$ at any given time $t$ can be expressed as a
biorthogonal density operator
\begin{eqnarray}\label{eqb-4}
    {\hat{\rho}}_{b}(t)=\sum_{n}p_n{\hat{\rho}}_{b,n}(t)=\sum_{n}p_n|\psi_n(t)\rangle\langle\widehat{\psi_n(t)}|,
\end{eqnarray}
where $p_n$ is the probability of the system being in the pure
state $|\psi_{n}(t)\rangle$, and
${\hat{\rho}}_{b,n}(t)=|\psi_n(t)\rangle{\langle\widehat{\psi_n(t)}|}$.

Let us first consider the system to be in the pure state
$|\psi_{n}(t)\rangle$.  When the eigenvalues of the
$\mathcal{PT}$-symmetric Hamiltonian $\hat{H}_{\mathcal{PT}}$ are
real numbers,  the system works in the $\mathcal{PT}$-symmetry
unbroken regime. In this case, according to eqs.~(\ref{eqa0-9})
and (\ref{eqa0-10}), one can obtain  the temporal evolution of the
density operator $\hat{\rho}_{b,n}(t)$,
\begin{eqnarray}
    &&\frac{d{\hat{\rho}}_{b,n}(t)}{dt}\cr
    &=&\frac{d{|\psi_n(t)\rangle\langle\widehat{\psi_n(t)}|}}{dt}\cr
    &=&\left( \frac{H_\mathcal{PT}}{i\hbar}|\psi_n(t)\rangle\langle\widehat{\psi_n(t)}|+
    |\psi_n(t)\rangle\langle\widehat{\psi_n(t)}|\frac{-H_\mathcal{PT}}{i\hbar}\right) \cr
    ~~~~~~~&=&\frac{1}{i\hbar}[H_\mathcal{PT}\hat{\rho}_{b,n}(t)-\hat{\rho}_{b,n}(t)H_\mathcal{PT}].\label{eqm-27a}
\end{eqnarray}
On the other hand, when the eigenvalues of the
$\mathcal{PT}$-symmetric Hamiltonian $\hat{H}_{\mathcal{PT}}$  are
imaginary numbers, the system  works in the
$\mathcal{PT}$-symmetry broken regime. In this situation,
according to eqs.~(\ref{eqa0-9}) and (\ref{eqa0-11}), one can find
that the temporal evolution of the density operator
$\hat{\rho}_{b,n}(t)$ follows
\begin{eqnarray}
    &&\frac{d{\hat{\rho}}_{b,n}(t)}{dt}\cr
    &=&\frac{d{|\psi_n(t)\rangle\langle\widehat{\psi_n(t)}|}}{dt}\cr
    &=&\left(\frac{H_\mathcal{PT}}{i\hbar}|\psi_n(t)\rangle\langle\widehat{\psi_n(t)}|+
    |\psi_n(t)\rangle\langle\widehat{\psi_n(t)}|\frac{H_\mathcal{PT}}{i\hbar}\right)\cr
    ~~~~~~~&=&\frac{1}{i\hbar}[H_\mathcal{PT}\hat{\rho}_{b,n}(t)+\hat{\rho}_{b,n}(t)H_\mathcal{PT}].\label{eqm-27b}
\end{eqnarray}
Moreover, one can verify  that
${\hat{\rho}_{b,n}}(t)={{U}_{\mathcal{PT}}(t)\hat{\rho}_{b,n}(0){U}_{\mathcal{PT}}^{'}(t)}$
satisfies the following relation
\begin{eqnarray}
    &&\frac{d{\hat{\rho}}_{b,n}(t)}{dt}\cr
    &=&\frac{d{U}_{\mathcal{PT}}(t)}{dt}\hat{\rho}_{b,n}(0){U}_{\mathcal{PT}}^{'}(t)
    +{U}_{\mathcal{PT}}(t)\hat{\rho}_{b,n}(0)\frac{d{U}_{\mathcal{PT}}^{'}(t)}{dt}\cr
    ~~~~~~~~~&=&\frac{1}{i\hbar}[H_\mathcal{PT}\hat{\rho}_{b,n}(t)-\hat{\rho}_{b,n}(t)H_\mathcal{PT}],\label{eqm-28a}
\end{eqnarray}
where
$\hat{U}_{\mathcal{PT}}(t)$=exp($-i\hat{H}_{\mathcal{PT}}t/\hbar$)
and
$\hat{U}^{'}_{\mathcal{PT}}(t)$=exp($i\hat{H}_{\mathcal{PT}}t/\hbar$)
are  time-evolution operators. Then, comparing eq.~(\ref{eqm-27a})
with eq.~(\ref{eqm-28a}), one can see  that
${\hat{\rho}_{b,n}}(t)={{U}_{\mathcal{PT}}(t)\hat{\rho}_{b,n}(0){U}_{\mathcal{PT}}^{'}(t)}$
is the general solution of eq.~(\ref{eqm-27a}) in the
$\mathcal{PT}$-symmetry unbroken regime. Similarly, it is easy to
prove that
${\hat{\rho}_{b,n}}(t)={{U}_{\mathcal{PT}}(t)\hat{\rho}_{b,n}(0){U}_{\mathcal{PT}}(t)}$
satisfies the following relation
\begin{eqnarray}
    &&\frac{d{\hat{\rho}}_{b,n}(t)}{dt}\cr
    &=&\frac{d{U}_{\mathcal{PT}}(t)}{dt}\hat{\rho}_{b,n}(0){U}_{\mathcal{PT}}(t)
    +{U}_{\mathcal{PT}}(t)\hat{\rho}_{b,n}(0)\frac{d{U}_{\mathcal{PT}}(t)}{dt}\cr
    &=&\frac{1}{i\hbar}[H_\mathcal{PT}\hat{\rho}_{b,n}(t)+\hat{\rho}_{b,n}(t)H_\mathcal{PT}].\label{eqm-28b}
\end{eqnarray}
One then has that
${\hat{\rho}_{b,n}}(t)={{U}_{\mathcal{PT}}(t)\hat{\rho}_{b,n}(0){U}_{\mathcal{PT}}(t)}$
is the general solution of eq.~(\ref{eqm-27b}) in the
$\mathcal{PT}$-symmetry broken regime by comparing
eq.~(\ref{eqm-27b}) with eq.~(\ref{eqm-28b}).

Let us now consider  the system to be in the mixed state
$\hat{\rho}_{b}(t)$. After substituting
${\hat{\rho}_{b,n}}(t)={{U}_{\mathcal{PT}}(t)\hat{\rho}_{b,n}(0){U}_{\mathcal{PT}}^{'}(t)}$
and
${\hat{\rho}_{b,n}}(t)={{U}_{\mathcal{PT}}(t)\hat{\rho}_{b,n}(0){U}_{\mathcal{PT}}(t)}$
into eq.~(\ref{eqb-4}), it is then straightforward that the
temporal evolution of the density operator $\hat{\rho}_{b}(t)$
follows
\begin{eqnarray}
    {\hat{\rho}}_{b}(t)={\hat{U}_{\mathcal{PT}}(t){\hat{\rho}}_{b}(0)\hat{U}_{\mathcal{PT}}^{'}(t)}\label{eqm-29a},\\
    {\hat{\rho}}_{b}(t)={\hat{U}_{\mathcal{PT}}(t){\hat{\rho}}_{b}(0)\hat{U}_{\mathcal{PT}}(t)}\label{eqm-29b},
\end{eqnarray}
where eq.~(\ref{eqm-29a}) corresponds to the case when the system
works in the $\mathcal{PT}$-symmetry unbroken regime, while
eq.~(\ref{eqm-29b}) corresponds to the case when the system works
in the $\mathcal{PT}$-symmetry broken regime. One can see that
eq.~(\ref{eqm-29a}) is eq.~(12) in the main text, while
eq.~(\ref{eqm-29b}) is eq.~(13) in the main text.


\begin{thebibliography}{1}

\bibitem{Symmetry1} A. Altland and M. R. Zirnbauer, ``Nonstandard Symmetry Classes in Mesoscopic Normal-Superconducting Hybrid Structures,'' Phys. Rev. B \textbf{55}, 1142-1161 (1997).
\bibitem{chiralsymmetry1} S. Malzard, C. Poli, and H. Schomerus, ``Topologically Protected Defect States in Open Photonic Systems with Non-Hermitian Charge-Conjugation and Parity-Time Symmetry,'' Phys. Rev. Lett. \textbf{115}, 200402 (2015).
\bibitem{chiralsymmetry3} Z. P. Gong, Y. Ashida, K. Kawabata, K. Takasan, S. Higashikawa, and M. Ueda, ``Topological Phases of Non-Hermitian Systems,'' {Phys. Rev. X} \textbf{8}, 031079 (2018).
\bibitem{chiralsymmetry2} K. Kawabata, K. Shiozaki, M. Ueda, and M. Sato, ``Symmetry and Topology in Non-Hermitian Physics,'' {Phys. Rev. X} \textbf{9}, 041015 (2019).
\bibitem{chiralsymmetry4}K. Y. Bliokh, J. Dressel, F. Nori, ``Conservation of the spin and orbital angular momenta in electromagnetism,'' {New J. Phys.} \textbf{16}, 093037 (2014).
\bibitem{topological2}  M. Li, X. Ni,  M. Weiner, A.  Al\`{u}, and  A. B. Khanikaev,  ``Topological phases and nonreciprocal edge states in non-Hermitian Floquet insulators,'' Phys. Rev. B \textbf{100}, 045423 (2019).


\bibitem{Noether1}  A. E. Noether, ``Invariante variations probleme,'' {Kgl Ges Wiss Nachr G\"{o}ttingen,'' Math. Phys. KI}\textbf{2}, 235-257 (1918).


\bibitem{Noether2}   N.  Ma, Y. Z. You and Z. Y. Meng, ``Role of Noether's Theorem at the Deconfined Quantum Critical Point,'' {Phys. Rev. Lett.} \textbf{122}, 175701 (2019).
\bibitem{Ehrenfest1} R. Shankar,  ``Principles of Quantum Mechanics,'' 2nd ed. (Springer, New York, 1994), https://doi.org/10.1007/978-1- 4757-0576-8.
\bibitem{Ehrenfest2} I. Marvian,  R. W. Spekkens, ``Extending Noether's theorem by quantifying the asymmetry of quantum states,'' {Nat. Commun.}  \textbf{5}, 3821 (2014).
\bibitem{Ehrenfest3} P. M. Zhang,  M. Elbistan,  P. A. Horvathy, P. Kosi\'{n}ski, ``A generalized Noether theorem for scaling symmetry,'' {Eur. Phys. J. Plus}  \textbf{135}, 223 (2020).


\bibitem{Noetheranalysis} K. Y. Bliokh, A. Y. Bekshaev, F. Nori, ``Dual electromagnetism: helicity, spin, momentum, and angular momentum,'' {New J. Phys.} \textbf{15}, 033026 (2013).

\bibitem{Noethercurrents} L. Burns, K. Y. Bliokh, F. Nori, J. Dressel, ``Acoustic versus electromagnetic field theory: scalar, vector, spinor representations and the emergence of acoustic spin,'' {New J. Phys.}  \textbf{22}, 053050 (2020).


\bibitem{Ehrenfest}  J. J. Garc\'{i}a-Ripoll,  V. M. P\'{e}rez-Garc\'{i}a, and  V. Vekslerchik, ``Construction of exact solutions by spatial translations in inhomogeneous nonlinear Schr\"{o}dinger equations,'' {Phys. Rev. E} \textbf{64}, 056602 (2001).


\bibitem{open1} Q. C. Wu, Y. H. Zhou, B. L. Ye, T. Liu and C. P. Yang, ``Nonadiabatic quantum state engineering by time-dependent decoherence-free subspaces in open quantum systems,'' {New J. Phys.} \textbf{23}, 113005 (2021).

\bibitem{r2} D. L. Li and C. Zheng, ``Non-Hermitian Generalization of R\'{e}nyi Entropy,''  Entropy,  \textbf{24}, 1563 (2022).
\bibitem{r5} X. E. Gao, D. L. Li, Z. H. Liu, and C. Zheng, ``Recent progress in quantum simulation of non-Hermitian,'' Acta Physica Sinica,  \textbf{71}, 240303 (2022).


\bibitem{conservationlaws2} H. Ramezani, T. Kottos, R. El-Ganainy, and D. N. Christodoulides, ``Unidirectional nonlinear PT-symmetric optical structures,'' {Phys. Rev. A} \textbf{82}, 043803 (2010).
\bibitem{interwining1} A. Mostafazadeh, ``Pseudo-Hermiticity versus $\mathcal{PT}$ symmetry: The necessary condition for the reality of the spectrum of a non-Hermitian hamiltonian,'' J. Math. Phys. \textbf{43}, 205-214  (2002).
\bibitem{interwining2} A. Mostafazadeh, ``Exact $\mathcal{PT}$-symmetry is equivalent to Hermiticity,'' J. Phys. A: Math. Gen. \textbf{36}, 7081-7091  (2003).
\bibitem{SRM} M. V. Berry, ``Optical lattices with $\mathcal{PT}$ symmetry are not transparent,'' J. Phys. A: Math. Theor. \textbf{41}, 244007  (2008).
\bibitem{SDM1} A. Mostafazadeh, ``Pseudo-Hermitian representation of quantum mechanics,'' Int. J. Geom. Methods Mod. Phys. \textbf{07}, 1191-1306  (2010).
\bibitem{SDM2} F. Ruzicka, K. S. Agarwal, Y. N. Joglekar, ``Conserved quantities, exceptional points, and antilinear symmetries in non-Hermitian systems,'' J. Phys.: Conf. Ser. \textbf{2038}, 012021  (2021).
\bibitem{RCM} Z. Bian, L. Xiao, K. Wang, X. Zhan, F. A. Onanga, F. Ruzicka, W. Yi, Y. N. Joglekar, and P. Xue, ``Conserved quantities in parity-time symmetric systems,'' Phys. Rev. Research \textbf{2}, 022039 (2020).
\bibitem{SPA} M. H. Teimourpour, R. El-Ganainy, A. Eisfeld, A. Szameit, and D. N. Christodoulides, ``Light transport in $\mathcal{PT}$-invariant photonic structures with hidden symmetries,'' Phys. Rev. A \textbf{90}, 053817 (2014).




\bibitem{Ehrenfest4}  J. D. H.  Rivero and L. Ge, ``Pseudochirality: A Manifestation of Noether's Theorem in Non-Hermitian Systems,'' {Phys. Rev. Lett.} \textbf{125}, 083902 (2020).



\bibitem{Non-Hermitian1}  C. M. Bender and S. Boettcher, ``Real Spectra in Non-Hermitian Hamiltonians Having $\mathcal{PT}$ Symmetry,''  {Phys. Rev. Lett.} \textbf{80},  5243-5246 (1998).
\bibitem{Non-Hermitian2} L. Ge, Y. D. Chong, and A. D. Stone, ``Conservation relations and anisotropic transmission resonances in one-dimensional $\mathcal{PT}$-symmetric photonic heterostructures,'' {Phys. Rev. A} \textbf{85}, 023802 (2012).

\bibitem{Non-Hermitianadd} B. Peng, S. K. Ozdemir, F. C. Lei, F. Monifi, M. Gianfreda, G. L. Long, S. H. Fan, F. Nori, C. M. Bender and L. Yang, ``Parity-time-symmetric whispering-gallery microcavities,'' Nat. Phys. \textbf{10}, 394-398 (2014).


\bibitem{Non-Hermitian3} H. Jing, S.K. Ozdemir, X. Y. Lu, J. Zhang, L. Yang, and F. Nori, ``$\mathcal{PT}$-Symmetric Phonon Laser,'' Phys. Rev. Lett. \textbf{113}, 053604 (2014).
\bibitem{Non-Hermitian4} V. V. Konotop, J. Yang, and D. A. Zezyulin,  ``Nonlinear waves in $\mathcal{PT}$-symmetric systems,'' {Rev. Mod. Phys.} \textbf{88}, 035002 (2016).
\bibitem{Non-Hermitian5} R. E. Ganainy, K. G. Makris, M. Khajavikhan, Z. H. Musslimani, and D. N. Christodoulides, ``Non-Hermitian physics and $\mathcal{PT}$ symmetry,'' {Nat. Phys.} \textbf{14}, 11-19 (2018).

\bibitem{r1} O. Sigwarth and  M. Christian, ``Time reversal and reciprocity,'' AAPPS Bull. \textbf{32}, 23 (2022).
\bibitem{r6} C. Zheng, ``Quantum simulation of $\mathcal{PT}$-arbitrary-phase-symmetric systems,'' EPL, \textbf{136}, 30002 (2021).


\bibitem{Non-Hermitian6} G. Q. Zhang, Z. Chen, Da Xu, N. Shammah, M. Liao, T.F. Li, L. Tong, S.Y. Zhu, F. Nori, and J.Q. You, ``Exceptional Point and Cross-Relaxation Effect in a Hybrid Quantum System,'' PRX Quantum \textbf{2}, 020307 (2021).
\bibitem{topological1} X. Ni,  D. Smirnova,  A. Poddubny, D. Leykam,  Y. Chong, and  A. B. Khanikaev,  ``$\mathcal{PT}$ phase transitions of edge states at $\mathcal{PT}$ symmetric interfaces in non-Hermitian topological insulators,'' Phys. Rev. B \textbf{98}, 165129 (2018).

\bibitem{optomechanics1} D. X. Chen, Y. Zhang, J. L. Zhao, Q. C. Wu, Y. L. Fang, C. P. Yang, and F. Nori, ``Quantum state discrimination in a $\mathcal{PT}$-symmetric system,'' {Phys. Rev. A} \textbf{106}, 022438 (2022).


\bibitem{optomechanics2}  H. Xu, D. G. Lai, Y. B. Qian, B. P. Hou, A. Miranowicz, and F. Nori, ``Optomechanical dynamics in the $\mathcal{PT}$- and broken-$\mathcal{PT}$-symmetric regimes,'' {Phys. Rev. A} \textbf{104}, 053518 (2021).

\bibitem{photonics1}  J. S. Tang, Y. T. Wang, S. Yu, D. Y. He, J. S. Xu, B. H. Liu, G. Chen, Y. N. Sun, K. Sun, Y. J. Han, C. F. Li, and G. C. Guo, ``Experimental investigation of the no-signalling principle in parity-time symmetric theory using an open quantum system,'' {Nat. Photonics} \textbf{10}, 642-646 (2016).
\bibitem{photonics2} Y. T. Wang, Z. P. Li, S. Yu, Z. J. Ke, W. Liu, Y. Meng, Y. Z. Yang, J. S. Tang, C. F. Li, and G. C. Guo, ``Experimental Investigation of State Distinguishability in Parity-Time Symmetric Quantum Dynamics,'' Phys. Rev. Lett. \textbf{124}, 230402 (2020).

\bibitem{microwave1}  J. Li, A. K. Harter, J. Liu, L. de Melo, Y. N. Joglekar, and L. Luo, ``Observation of parity-time symmetry breaking transitions in a dissipative Floquet system of ultracold atoms,'' {Nat. Commun.} \textbf{10}, 855 (2019).

\bibitem{microwave2}   H. Z. Chen, T. Liu, H. Y. Luan, R. J. Liu, X. Y. Wang, X. F. Zhu, Y. B. Li, Z. M. Gu, S. J. Liang, H. Gao,  L. Lu, L. Ge, S. Zhang, J. Zhu, and R. M. Ma, ``Revealing the missing dimension at an exceptional point,'' {Nat. Phys.} \textbf{16}, 571-578 (2020).


\bibitem{r3} C. Wu, A. Fan, and S. D. Liang, ``Complex Berry curvature and complex energy band structures in non-Hermitian graphene model,'' AAPPS Bull.  \textbf{32}, 39 (2022).
\bibitem{r4} C. Zheng, L. Hao, and G. L. Long, ``Observation of a fast evolution in a parity-time-symmetric system,'' Phil. Trans. R. Soc. A.  \textbf{371}, 20120053 (2013).



\bibitem{CriticalPhenomena1} L. Xiao, K. K. Wang, X. Zhan, Z. H. Bian, K. Kawabata, M. Ueda, W. Yi, and P. Xue, ``Observation of Critical Phenomena in Parity-Time-Symmetric Quantum Dynamics,'' {Phys. Rev. Lett.} \textbf{123}, 230401 (2019).
\bibitem{CriticalPhenomena2}  Y. Ashida, S. Furukawa, and M. Ueda, ``Parity-time-symmetric quantum critical phenomena,'' {Nat. Commun.} \textbf{8}, 15791 (2017).

\bibitem{energytransfer1}  H. Xu, D. Mason, L. Jiang, and J. G. E. Harris, ``Topological energy transfer in an optomechanical system with exceptional points,'' {Nature} \textbf{537}, 80-83 (2016).
\bibitem{energytransfer2}  J. Doppler, A. A. Mailybaev, J. B\"{o}hm, U. Kuhl, A. Girschik, F. Libisch, T. J. Milburn, P. Rabl, N. Moiseyev, and S. Rotter, ``Dynamically encircling an exceptional point for asymmetric mode switching,'' {Nature} \textbf{537}, 76-79 (2016).

\bibitem{InformationRetrieva1}  K. Kawabata, Y. Ashida, and M. Ueda, ``Information Retrieval and Criticality in Parity-Time-Symmetric Systems,'' {Phys. Rev. Lett.} \textbf{119}, 190401 (2017).
\bibitem{InformationRetrieva2}  J. W. Wen, C. Zheng, Z. D. Ye, T. Xin, and G. L. Long, ``Stable states with nonzero entropy under broken $\mathcal{PT}$ symmetry,'' {Phys. Rev. Research} \textbf{3}, 013256 (2021).
\bibitem{InformationRetrieva3}  Y. L. Fang, J. L. Zhao, Y. Zhang, D. X. Chen, Q. C. Wu, Y. H. Zhou, C. P. Yang, and F. Nori, ``Experimental demonstration of coherence flow in $\mathcal{PT}$- and anti-$\mathcal{PT}$-symmetric systems,'' {Commun. Phys.} \textbf{4}, 223 (2021).
\bibitem{invariant1} H. Shen, B. Zhen, and  L. Fu,  ``Topological band theory for non-Hermitian Hamiltonians,'' Phys. Rev. Lett. \textbf{120}, 146402 (2018).
\bibitem{invariant2} K. Chen and A. B. Khanikaev,  ``Non-Hermitian $C_{\textrm{NH}}$=2 Chern insulator protected by generalized rotational symmetry,'' Phys. Rev. B \textbf{105}, L081112 (2022).
\bibitem{APT} Y. Choi, C. Hahn, J. W. Yoon, and S. H. Song, ``Observation of an anti-$\mathcal{PT}$-symmetric exceptional point and energy-difference conserving dynamics in electrical circuit resonators,'' Nat. Commun. \textbf{9}, 2182 (2018).



\bibitem{Biorthogonal1} D. C. Brody, ``Biorthogonal quantum mechanics, J. Phys. A-Math. Theor. \textbf{47}, 035305 (2013).
\bibitem{Biorthogonal2} F. K. Kunst, E. Edvardsson, J. C. Budich, and E. J. Bergholtz, ``Biorthogonal bulk-boundary correspondence in non-Hermitian systems,'' {Phys. Rev. Lett.} \textbf{121}, 026808 (2018).
\bibitem{Biorthogonal3} Q. C. Wu, Y. H. Chen, B. H. Huang, Y. Xia, and J. Song, ``Reverse engineering of a nonlossy adiabatic Hamiltonian for non-Hermitian systems,'' {Phys. Rev. A} \textbf{94}, 053421 (2016).
\bibitem{Biorthogonal4}C. Y. Ju, A. Miranowicz, G.Y. Chen, F. Nori, ``Non-Hermitian Hamiltonians and no-go theorems in quantum information,'' Phys. Rev. A \textbf{100}, 062118 (2019).


\bibitem{masking1} K. Modi,  A. K. Pati,  A. Sen, U. Sen, ``Masking quantum information is impossible,'' {Phys. Rev. Lett.} \textbf{120}, 230501 (2018).
\bibitem{masking2} Z. H. Liu,  X. B. Liang,  K. Sun,  Q. Li,  Y. Meng,  M. Yang, B. Li, J. L. Chen, J. S. Xu, C. F. Li and  G. C. Guo, ``Photonic implementation of quantum information masking,'' {Phys. Rev. Lett.} \textbf{126}, 170505 (2021).


\bibitem{non-Hermitianadd} C.Y. Ju, A. Miranowicz, F. Minganti, C. T. Chan, G. Y. Chen, and F. Nori, ``Einstein's quantum elevator: Hermitization of non-Hermitian Hamiltonians via a generalized vielbein formalism,'' Phys. Rev. Research \textbf{4}, 023070 (2022).

\bibitem{rho1}D. C. Brody and E. M. Graefe, ``Mixed-state Evolution in the Presence of Gain and Loss,'' Phys. Rev. Lett. \textbf{109}, 230405 (2012).
\bibitem{rho2} L. Xiao, X. Zhan, Z. H. Bian, K. K. Wang, X. Zhang, X. P. Wang, J. Li, K. Mochizuki, D. Kim, N. Kawakami, W. Yi, H. Obute, B. C. Sanders, and P. Xue, ``Observation of topological edge states in parity-time-symmetric quantum walks,'' Nat. Phys. \textbf{13}, 1117-1123 (2017).

\bibitem{Tomography}D. F. V. James, P. G. Kwiat, W. J. Munro, and A. G. White, ``Measurement of qubits,'' Phys. Rev. A \textbf{64}, 052312 (2001).
\bibitem{Tomography1} M. Naghiloo, M. Abbasi, Y. N. Joglekar, and K. W. Murch, ``Quantum state tomography across the exceptional point in a single dissipative qubit,'' Nat. Phys. \textbf{15}, 1232 (2019).


\end{thebibliography}
\end{document}